\documentclass{article}

\usepackage[english]{babel}

\usepackage[super,sort&compress,comma]{natbib} 
\usepackage[version=3]{mhchem}
\usepackage[a4paper, left=2cm, right=2cm, top=1.785cm, bottom=2.0cm]{geometry}

\usepackage{subcaption}
\usepackage{abstract}
\usepackage{amsmath}
\usepackage{titling}
\usepackage{authblk}
\usepackage{amssymb}
\usepackage{balance}
\usepackage{mathptmx}
\usepackage{sectsty}
\usepackage{graphicx} 
\usepackage{tabularx} 
\usepackage{multirow}
\usepackage{makecell}
\usepackage[colorlinks=true, allcolors=blue]{hyperref}
\usepackage{xr}
\usepackage{hhline}

\renewcommand\floatpagefraction{.9}

\begin{document}
	\title{From oligomers to entangled polymers: How to train a transferable machine learning interatomic potential}
	
	\author[a]{\normalsize Mirko Fischer}
	\author[a]{\normalsize Andreas Heuer}
	
	\affil[a]{Institute of Physical Chemistry, University of Münster, Corrensstraße 28/30, 48149 Münster, Germany}
	\date{}
	
	\maketitle
	\begin{abstract}
		Over the past decade, Machine Learning Interatomic Potentials (MLIPs) have emerged as a powerful technique for performing molecular dynamics (MD) simulations with nearly ab initio accuracy. Alongside the development of new descriptors and advanced machine learning architectures, sophisticated procedures for the generation of diverse and accurate reference datasets have been established. To date, research has focused primarily on MLIPs for crystalline or amorphous inorganic and small molecular systems; however, large macromolecules such as polymers remain underrepresented in the literature, despite beeing an important class of materials. In this work, we investigate several aspects of developing MLIPs for polymers, utilizing polyethylene as a representative, yet simple model system.  
		
		First, we compare various local atomic descriptors, identifying the Atomic Cluster Expansion (ACE) as the most effective for this application. Second, we implement an automatized active learning scheme to efficiently generate diverse training data and demonstrate that ACE potentials fitted on small oligomers are transferable to larger polymers. Given that the accurate reproduction of the density depends critically on a correct description of intermolecular interactions, which are far more complex to learn than intramolecular interactions, we carefully evaluate the performance of the ACE potentials with respect to non-bonded interactions.
		By utilizing the computationally efficient OPLS-AA force field as a ground truth reference, we are able to perform a direct comparison of nanosecond-scale MD trajectories resulting from the ACE and reference potential. We find that the ACE potential accurately reproduces key thermodynamic, structural and dynamical properties.
	\end{abstract}
	\section{Introduction}
	
	Polymers represent an important class of materials and are used for various different applications, including but not limited to packaging \cite{Barlow.2013, Teixeira.2025}, medicine \cite{Pack.2005, Torchilin.2001} or battery electrolytes \cite{Long.2016, Zhou.2019, Hallinan.2013}. Their properties are determined by the chemical composition, shape of individual polymer molecules and also their processing. \cite{Krevelen.2009, Strobl.2010, Fetters.1994}
	The design of new or tuning of known polymeric systems is therefore highly non-trivial and requires deep expertise, laboratory experiments as well as computational studies. \cite{Sattari.2021, Askadskii.2003} 
	For years, polymers have been studied in computational chemistry and physics to understand their basic underlying physics. Whereas ab initio methods like Density Functional Theory (DFT) can provide an understanding of polymerization processes and electronic properties, it is limited to small systems of a couple of hundred atoms and time scales of picoseconds. Therefore, large polymeric systems can not be modeled. \cite{Yurtsever.2002, Salzner.2014}
	Contrary to DFT, molecular dynamics (MD) simulations are unable to describe the polymerization process and also less accurate but enable simulations of large bulk systems with thousands of atoms at the nanosecond timescale, providing insights into the polymeric structure, coiling behaviour and size as well as entanglements. Additionally, also the dynamics can be characterized, which might be strongly governed by the underlying structure, e.g. by entanglements. \cite{Litvinov.2013, Stepisnik.2016,Tejedor.2023}
	Typically, this requires nanoseconds long simulations due to slow relaxation processes and restricted dynamics. \cite{Cates.1987, Kremer.1990}
	
	Recently, Machine Learning Interatomic Potentials (MLIPs) have evolved as a powerful tool, which aim to combine the efficiency of classical MD simulations with the accuracy of ab initio calculations. Whereas classical MD relies on handcrafted, analytical and physical interpretable potentials, MLIPs rely on flexible machine learning models, which allow to tune their accuracy based on the number of parameters and the accuracy of the underlying training data. \cite{Deringer.2019, Zuo.2020, Jacobs.2025} Typically, they predict energie and forces based on the local atomistic environment, which can be represented via atomic descriptors like SOAP \cite{Bartok.2013, De.2016}, LMBTR \cite{Huo.2022} or ACE \cite{Drautz.2019}.
	Another advantage of MLIPs is that no fixed topology of molecules needs to be defined, allowing in principle reactions to take place. The MLIPs can learn the molecular topology directly from the data.
	This is not possible with classical potentials, for which the topology needs to be fixed to describe bonded interactions via analytical functions.
	
	So far, MLIPs have been successfully trained and applied for various crystalline and amorphous systems as well as small organic molecules, including the description of reactions. \cite{Deringer.2019, Kulichenko.2024} On the one hand, there are highly specialized MLIPs, trained specifically on a particular system of interest and on the other hand, there are so-called foundation models, which are trained on large data sets comprising a huge portion of the chemical space \cite{Batatia.2025, Allen.2024, Li.2026}.  
	
	For macromolecules like polymers to our knowledge there exists only a handful of studies in which MLIPs for polymers are trained, applied and evaluated. Foundation models for polymers do not exist up to know, although recently with OPoly26 a large data set was published which may serve as a starting point \cite{Levine.2025}. 
	
	Potential future applications of MLIPs for polymers would allow for example to study polymerization reactions and electronic properties under operating conditions, which might have a strong impact on structure and dynamics. 
	
	The existing studies focus mainly only on certain aspects of MLIPs of polymers, but a systematic evaluation is missing. Long \textit{et al.} provide an overview about the recent developments and future challenges \cite{Long.2024}. One early study focused on crystalline organic polymers, demonstrating the importance of correctly fitted van der Waals interactions \cite{Hong.2021}.
	For polyethylene glycole (PEG), it was demonstrated that using a GAP potential with single oligomer structures as training data, correct experimental densities and stable simulations for a single long polymer chain can be achieved \cite{vanderOord.2023}. Another work on PEG employed clusters of small oligomers as training data and also characterized the local structure and density, diffusivity and heat capacity depending on the oligomer size. The results were compared to the OPLS4 force field \cite{Mohanty.2023}. For Polyacrylonitrile (PAN) structures obtained with a neural network potential were compared to structures obtained with the ReaxFF force field \cite{Chahal.2024}. 
	All these studies trained MLIPs based on oligomers that were much smaller than the polymers to which they were applied, indicating that this approach is effective. Except of the last study, all others used also some active learning scheme to efficiently sample training data.
	Other approaches aim to refit analytical potentials using ML techniques or replace only some interactions by ML potentials \cite{Chen.2024, Wang.2021}.
	
	Whereas different aspects of fitting MLIPs for polymers are distributed across these different publications a systematic investigation is missing so far. It is well established that polymer structures can be reproduced well, however, for the dynamics this is still open to discussion, since reference simulations with DFT for polymers are not feasible on the necessary time scales. Based on the existing literature, we identified several key questions, which remain unanswered so far:
	(1) Which local atomic descriptor is best suited for polymeric systems?
	(2) Do MLIPs trained on single oligomer molecules perform better than those trained on clusters, and what is the optimal cluster size?
	(3) What is the minimum oligomer size required during training to maintain a high accuracy when applying the MLIP to a polymeric system?
	(4) Can a carefully trained MLIP preserve the polymer topology during nanosecond-scale simulations?
	(5) How do small prediction errors affect key structural and dynamical properties in oligomers and polymers?
	(6) Can the dynamics of large entangled polymers be accurately reproduced by an MLIP?
	
	Question (1) and (2) are unexplored so far, since a comparison of different descriptors and of different cluster sizes as training data for the same system are missing in the existing literature. Both questions are explicitly considered in this work.  
	
	The issue of minimum chain length (3) was recently explored by Hooven \textit{et al.} \cite{Hooven.2026} through the systematic training of MACE potentials on a series of single alkane molecules from methane to octane. While they demonstrated that potentials trained on butane or larger species are highly transferable with regards to energy and force predictions, the performance of these models in bulk phases remains untested.
	
	We aim to fill this gap by training MLIPs for alkanes and polyethylene as a simple model system and directly apply them for bulk phase simulations. Thereby, we furthermore answer questions (4), (5) and (6) which were to our knowledge not addressed so far in existing literature. 
	
	This work is structured as follows: First, we evaluate various local atomic descriptors assessing their similarity across varying alkane and cluster sizes. We then systematically train ACE potentials on clusters of alkanes with different lengths, utilizing an active learning scheme to efficiently sample the relevant configurational space. 
	This training process is analyzed to determine the optimal chain length and cluster size, with a particular emphasis on the accurate reproduction of intermolecular interactions. The resulting optimal ACE potential is then applied to bulk simulations of polymers of varying lengths to analyze their structural and dynamical properties, which are benchmarked against reference simulations.
	
	Because these fundamental methodological questions can be addressed independently of the accuracy of the training data, we utilize the computationally efficient OPLS-AA force field \cite{Jorgensen.1996} as ground truth reference instead of computationally costly DFT. This approach enables a rapid sampling of the configurational space and therefore enables the testing of various active learning configurations and data sets. Furthermore it enables nanosecond-scale reference simulations, which are essential to characterize the inherently slow dynamics of polymer chains.
	
	\section{Methods}
	
	\subsection{Polymer model systems}
	Polyethylene (PE) was selected as a representative model system due to its simplicity, comprising only carbon and hydrogen as atomic species.
	In MD simulations, PE exhibits only small long-range electrostatic interactions due to its low partial charges.
	
	These characteristics make PE the ideal candidate for rapid training and also evaluation of MLIPs, which rely typically on the locality assumption. However, despite the structural simplicity of PE, the developement of an accurate MLIP remains a significant challenge. Specifically, very weak van-der-Waals interactions, which are orders of magnitude smaller than intramolecular interactions, are often poorly captured by MLIPs. Previous research has indicated that carefully curated reference data, including single molecules in vacuum and volume scans, are necessary to overcome the inherent difficulty MLIPs face when predicting energies and forces across several orders of magnitudes \cite{Magdau.2023}. %EC_EMC paper 
	In this paper, we examine PE chains of varying lengths, ranging from small alkanes (\ce{C2H6}) up to \ce{C512H1026} and train MLIPs either for individual chain lengths or combined sets. 
	3D molecular structures are generated from SMILES strings using RDKit \cite{Landrum.} up to \ce{C16H34} and an in-house tool beyond.
	
	\subsection{Atomic Cluster Expansion model}
	The atomic cluster expansion (ACE), originally proposed by Drautz \cite{Drautz.2019}, provides a generalized framework for descriptor-based MLIPs. It translates the local atomic environment of an atom into a vector-based descriptor, defined by a set of many-body basis functions. 
	Using this formalism an atomic property of atom $i$ can be expanded as
	\begin{equation}
		\phi_i^{(p)} = \sum_{\boldsymbol{\nu}}^{n_{\boldsymbol{\nu}}} c_{\boldsymbol{\nu}^{(p)}} \boldsymbol{B}_{i\boldsymbol{\nu}}
	\end{equation}
	
	where $\boldsymbol{B}_{i\boldsymbol{\nu}}$ are the basis functions and $c_{\boldsymbol{\nu}^{(p)}}$ are the corresponding coefficients which need to be fitted via linear regression. $\boldsymbol{\nu}=K\boldsymbol{nl}$ is a multi-index, where $K$ refers to the number of atoms considered in a multi-body interaction and $\boldsymbol{n}$ and $\boldsymbol{l}$ are quantum numbers of atomic orbitals, that appear in the construction of the basis functions $\boldsymbol{B}_{i\boldsymbol{\nu}}$.
	In general, the atomic energy is expressed as \begin{equation}
		E_i = F\left(\phi_i^{(1)}, \phi_i^{(1)}, ...,  \phi_i^{(P)}\right)
	\end{equation}
	However, in the simplest case of only one atomic property it is equal to the atomic energy. 
	\begin{equation}
		E_i = \phi_i^{(1)}
	\end{equation}
	
	Within this paper, we employ only one atomic property for simplicity. 
	Furthermore, we restrict the expansion to two-body, three-body and four-body basis functions. This matches the type of interactions encountered in the OPLS-AA force field \cite{Jorgensen.1996}which we employ as a reference (see Computational details). In the limit of an infinite number of basis functions of these body-orders, the OPLS-AA force field should be perfectly reproduced.
	However, for computational efficiency, we select only 606 basis functions in total. Furthermore, the cutoff radius to describe the local atomic environment is limited to $7\:\mathrm{\AA}$. 
	Note that many foundation models, fitted to large data bases, employ even smaller cutoff radii of $5\:\mathrm{\AA}$ or $6\:\mathrm{\AA}$ \cite{Riebesell.}.  
	
	The coefficients of the basis functions are optimized by minimizing a loss function, which is given by
	\begin{equation}
		L = (1-\kappa) \sum_{n=1}^{N_\mathrm{struct}} w_n^{(E)} \left(\frac{E_n^{\mathrm{ACE}}- E_n^{\mathrm{REF}}}{n_{at,n}}\right)^2 
		+ \kappa 
		\sum_{n=1}^{N_\mathrm{struct}} 
		\sum_{i=1}^{n_{at,n}}
		w_{ni}^{(F)} \left(\frac{F_n^{\mathrm{ACE}}- F_n^{\mathrm{REF}}}{n_{at,n}}\right)^2 
	\end{equation}
	
	where $E_n^{\mathrm{ACE}}$, $E_n^{\mathrm{ACE}}$,  $F_n^{\mathrm{ACE}}$ and $F_n^{\mathrm{REF}}$ are the energies and forces predicted by ACE and the corresponding ground truth. $N_\mathrm{struct}$ is the number of different structures used for training an ACE potential and $n_{at,n}$ is the number of atoms in structure $n$. $w_n^{(E)}$ and $w_{ni}^{(F)}$ are weighting factors which can be used to weight important structures higher than less important structures. $\kappa$ weights the energies and the forces in the loss function. Within this paper, we weight the forces comparatively high by setting $\kappa=0.95$, since they ultimately govern the dynamics, while an offset in the potential energy has no effect. 
	
	\subsection{Active Learning Workflow}
\begin{figure*}[h]
		\centering
		\includegraphics{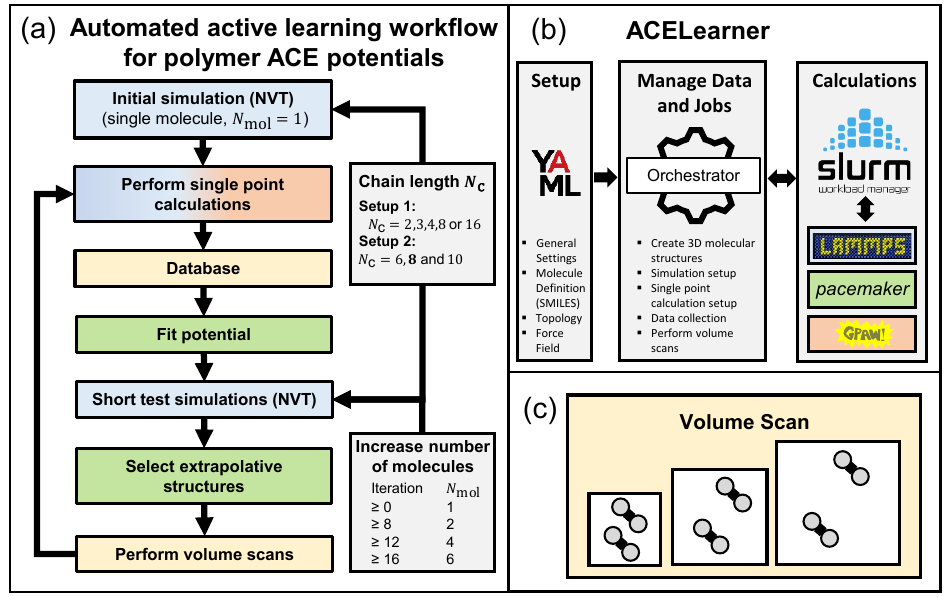}
		\caption{a) Schematic active learning workflow for fitting ACE potentials applied within this work. b) Illustration of ACELearner as an orchestrator tool to manage the active learning workflow based on YAML setup files. c) Visual explanation of volume scans. }
		\label{fig:al_workflow}
	\end{figure*}

	To enable the autonomous training of ACE potentials with minimal human intervention and computational overhead, we developed an active learning (AL) workflow that integrates existing software tools with custom-developed implementations. The full workflow is displayed in \autoref{fig:al_workflow}a and explained below.
	
	\subsubsection{Initialization}
	
	The AL procedure is initialized by performing MD simulations of a single molecule utilizing the LAMMPS software \cite{Thompson.2022}. Snapshots from these simulations serve as the initial structural data base for the subsequent potential fitting. 
	The reference energies and forces for these structures are calculated using the selected reference method. For DFT-based calculations the GPAW software \cite{Mortensen.2024} is integrated into the workflow, while LAMMPS is utilized for MD-based reference calculations. 
	
	\subsubsection{ACE potential fitting}
	
	An ACE potential is fitted to the initial structures using the \textit{pacemaker} software tool \cite{Bochkarev.2022}, employing the L-BFGS solver for a maximum of 1000 iterations. 
	Note that each potential is fitted from scratch during further AL iterations to avoid being stuck in local minima.
	More detailed setting of \textit{pacemaker} are described in the SI ( \autoref{si_sec:pacemaker_settings}).
	
	\subsubsection{Test simulations}
	
	To evaluate the ACE potential, several MD test simulations are initiated. The initial setup consists of a single molecule in a periodic box; however, as the active learning process evolves, additional molecules are introduced at predefined iterations to increase system complexity. 
	This strategy is designed to prioritize the accurate fitting of intramolecular interactions before subsequently incorporating intermolecular interactions through small molecular clusters and bulk systems. Within this work, at iteration 8, 12 and 16 further molecules are added, such that the setup consists of 2, 4 and 6 molecules. After 20 iterations, the active learning is stopped.
	
	Note that to ensure a more diverse exploration of different local atomic configurations, several short test simulations (typically 50 ps within this work) with different initial structures and velocities are performed rather than one long test simulation. This is particularly efficient during the first AL iterations, where only a few extrapolative structures per test simulation are encountered before the simulation is stopped due to a high extrapolation grade of the ACE potential (see below).
	
	\subsubsection{Extrapolation grade}
	In the ACE formalism, each local atomic environment $i$ is represented by a corresponding vector of basis functions $\boldsymbol{B_i}=(B_{i1}, B_{i2}, \dots, B_{in_\nu})$.
	Each vector can be also represented as a combination of a subset of vectors $\boldsymbol{B}_k^A$ that span the largest volume of at most $n_\nu$ linear independent vectors among all vectors representing atomic environments. \cite{Lysogorskiy.2023}
	\begin{equation}
		\boldsymbol{B}_i =
		\sum_{k=1}^{n_\nu} \gamma_k^{(i)}\boldsymbol{B}_k^A,
		\label{eq:gamma}
	\end{equation}
	The coefficients $\gamma_k^{(i)}$ can be identified as a metric to measure the extrapolation for a given atomic environment $i$.
	Together, the vectors $\boldsymbol{B}_k^A$ form the active set matrix:
	\begin{equation}
		\hat{A} = \left( \begin{matrix} 
			B_{11} & \dots & B_{1n_\nu} \\
			\vdots & \ddots & \vdots \\
			B_{n_\nu1} & \dots & B_{n_\nu n_\nu} \\
		\end{matrix}\right)
	\end{equation}
	By solving the matrix form of \autoref{eq:gamma} for $\gamma_k^{(i)}$, it can be directly calculated. The maximum occurring $\gamma_k^{(i)}$ for atom $i$ is then defined as the corresponding extrapolation grade for this atom and can be calculated as: \cite{Lysogorskiy.2023}
	\begin{equation}
		\gamma_i = \max_k \left(|\gamma_k^{(i)}|\right) = \max |\boldsymbol{B}_i \cdot \hat{A}^{-1}|
	\end{equation}
	
	If $|\gamma_i|<1$ there is only interpolation, however, if $|\gamma_i|>1$ extrapolation is observed. $|\gamma_i|=1$ means that a given atomic environment is exactly represented in the active set. 
	
	In practice, all structures with an extrapolation grade larger than $\gamma_\mathrm{ext}$, which is typically between 1-10, are collected. As a native choice, $\gamma_\mathrm{ext}=2$ was selected for this work. Furthermore, if $\gamma_i$ exceeds the threshold $\gamma_\mathrm{stop}$, the complete simulation is interrupted. Unless otherwise specified, a value of $\gamma_\mathrm{stop} = 3.5$ is used throughout this study.
	
	From all collected extrapolative structures, a subset of most diverse structures is selected for addition to the training data base, using the pacemaker utility tools \cite{Bochkarev.2022, Lysogorskiy.2023}, to avoid the addition of similar or duplicate structures. 
	
	\subsubsection{Volume scans}
	
	To further improve the potential with respect to non-bonded interactions, a volume scan is performed on some or all structures within this subset. During the volume scan the simulation box is scaled to increase the distance between individual molecules while preserving all intramolecular distances, as illustrated in \autoref{fig:al_workflow}b. This procedure enables the MLIP to more effectively differentiate between inter- and intramolecular interactions. \cite{Magdau.2023}
	If not mentioned otherwise, a subset of maximum 50 structures is selected within this work per AL iteration, and a volume scan is performed for not more than 25 of them, where for each structure 6 different volumes are selected by scaling the original box length with equidistant factors between 1.0 and 1.2.
	For all of those structures, reference calculations are performed prior to their addition to the data base and the subsequent fitting of the potential. 
	
	\subsubsection{Termination}
	
	If no extrapolated structures are identified, it is checked whether the maximum number of molecules in the test simulation has been reached. If not, additional molecules are introduced earlier than initially scheduled. 
	The AL procedure terminates either when the maximum number of iterations is reached or when no further extrapolated structures are identified for the maximum-sized system.
	
	\subsubsection{Refining the final potential}
	Upon completion of the AL workflow for a combined set of chain lengths $N_c=6,8\,\mathrm{and}\,10$, a refined potential was trained on the full dataset.
	Since the BFGS optimizer \cite{Fletcher.1987} employed in the \textit{pacemaker} tool is susceptible to local minima, five independent ACE potentials were trained using different random seeds to ensure a more robust optimization.
	The best performing potential was then selected as a starting point for further optimization. An additional 1000 L-BFGS iterations were performed using an adjusted loss function with $\kappa=0.05$ to ensure the accurate reproduction of energies and minimize any systematic energy offsets.
	95 \% of all structures collected during the AL process were used for training the refined potential and 5 \% were used for testing.
	A similar strategy was also applied to fit potentials to with the data sets obtained after iterations 8, 12 and 16 to evaluate the effect of the additional training data without any bias.
	
	\subsubsection{ACELearner software}
	
	All important parameters of the AL procedure are configurable via a YAML setup file. This file serves as the input for the ACELearner software, which acts as an orchestrator for the entire workflow (see \autoref{fig:al_workflow}c). 
	ACELearner automatically initializes the reference simulations from a SMILES string of the molecule under consideration, manages the AL procedure and data handling, and interfaces with the Slurm workload manager to execute computationally costly potential fitting and testing and all reference calculations in parallel within a high-performance-computing (HPC) environment.
	The detailed setup and parameters of the ACELearner AL workflow are described in the SI (\autoref{si_sec:al_settings}).
	
	\section{Results and Discussion}
	
	\subsection{Comparison of local descriptors}
	
	\begin{figure*}[h]
		\centering
		\includegraphics[width=\linewidth]{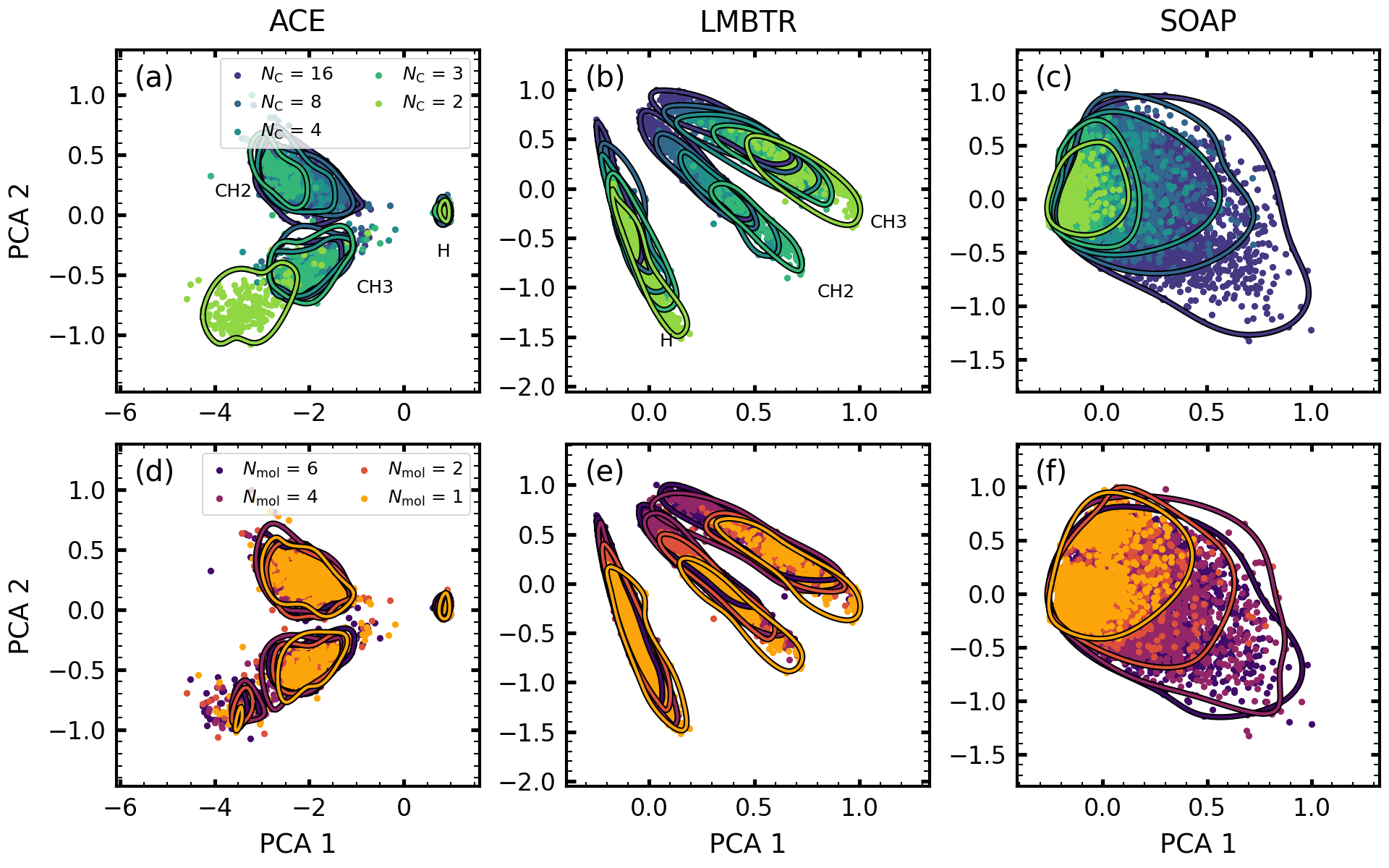}
		\caption{Comparison of regions within the descriptor space covered by clusters of different sizes $N_\mathrm{mol}$ and with different chain length $N_c$. a)-c) are color-coded by the chain length and d)-f) are color coded by the size of molecules within a cluster. As descriptors a), d) ACE, b), e) LMBTR and c), f) SOAP are compared. The concave hulls of the clusters for individual $N_c$ and $N_\mathrm{mol}$ are indicated as colored lines.}
		\label{fig:descriptor}
	\end{figure*}
	
	For reliable and accurate MD simulations with MLIPs, it is essential that the local atomic environments encountered during the MD simulation are well-represented within the training set. Otherwise, the MLIP may extrapolate which can lead to non-physical or unstable simulations. 
	The similarity of local atomic environments can be characterized by comparing the corresponding local atomic descriptors across different structures. In this work, we evaluate three descriptors, commonly employed in the development of MLIPs: The Atomic Cluster Expansion (ACE) descriptor, the Local Many-Body Tensor Representation (LMBTR) \cite{Huo.2022} and the Smooth Overlap of Atomic Positions (SOAP) \cite{Bartok.2013, De.2016}. 
	These descriptors were applied to a dataset comprising single molecules and clusters of up to six oligomers. The oligomer chain lengths vary from two carbon atoms (ethane) up to sixteen (hexadecane), with all oligomers within a cluster maintaining a consistent size. Details about the dataset can be found in the SI (\autoref{si_sec:dataset_descriptor}).
	
	\autoref{fig:descriptor} displays the two most important principle components (PCs) from a principle component analysis (PCA) \cite{Gewers.2022} of the high-dimensional descriptors. In \autoref{fig:descriptor}a-c, the local atomic environment described by ACE, LMBTR and SOAP are color-coded by the oligomer size. Three distinct observations can be made. 
	
	First, ACE and LMBTR effectively cluster atoms of different atom types, whereas SOAP fails to separate different atomic environments clearly. Furthermore, ACE and LMBTR also manage to distinguish between central (\ce{CH2}) and terminal (\ce{CH3}) carbon atoms, which is critical for a correct representation of the oligomer topology. If a descriptor fails to distinguish between \ce{CH2} and \ce{CH3}, there is a risk that oligomer chains may break or merge during simulations.
	
	Second, smaller oligomers occupy a smaller region of the ACE and SOAP descriptor space. As oligomer length increases, a larger region is explored. This is driven by two effects: an increased density of larger oligomer clusters due to enhanced van der Waals interactions, and the emergence of different structural motifs, such as increased flexibility and bending in longer chains. These effects are most pronounced in the SOAP descriptor, which is highly sensitive to the system density, and less significant in the ACE descriptor. For the LMBTR descriptor a shift of the occupied regions is observed with increasing chain length rather than an expansion. 
	
	Third, for ACE, we observe that the descriptor space for hydrogen and terminal carbon atoms is nearly fully explored using as small oligomers as propane. The only exception is the region occupied by terminal carbon atoms in ethane, which differs because the terminal carbon atoms are bonded directly to one another. 
	
	While the region of the descriptor space that describes the central carbon atom expands as the oligomer size increases from propane ($N_c=3$) to octane ($N_c=8$), the regions descried by octane and hexadecane ($N_c=16$) nearly fully overlap. Consequently, it is expected that octane is sufficient to accurately represent central carbon atoms in much larger oligomers or polymers. 
	
	In \autoref{fig:descriptor}d-f the local atomic environments are color-coded by cluster size. We observe effects similar to those seen when increasing the oligomer length.
	For SOAP, single molecules and small clusters occupy a limited region of the descriptor space while larger clusters occupy a significantly extended region. For LMBTR the regions are slightly expanded and additionally shifted. This expansion and shift diminish as the cluster size increases, becoming marginal when increasing the cluster size from four to six molecules.
	
	This suggests that for bulk simulation SOAP and LMBTR require the inclusion of larger clusters or, ideally, bulk structures in the training set of any MLIP to avoid extrapolation. In contrast, the ACE descriptor shows a similar occupation of the descriptor space by single molecules and larger clusters. This implies that an MLIP trained on a single molecule using ACE could potentially yield an accurate description of bulk structures.
	
	Consequently, while LMBTR and SOAP require training sets composed of large clusters or long-chain polymers to ensure a sufficient overlap with bulk structures, the ACE descriptor remains robust even when trained only on small clusters or single molecules of relatively short oligomers such as octane. Based on these findings, we have selected the ACE descriptor as the ideal candidate for our subsequent studies.
	
	\subsection{Active Learning}
	
	\begin{figure*}[h]
		\centering
		\includegraphics[width=\linewidth]{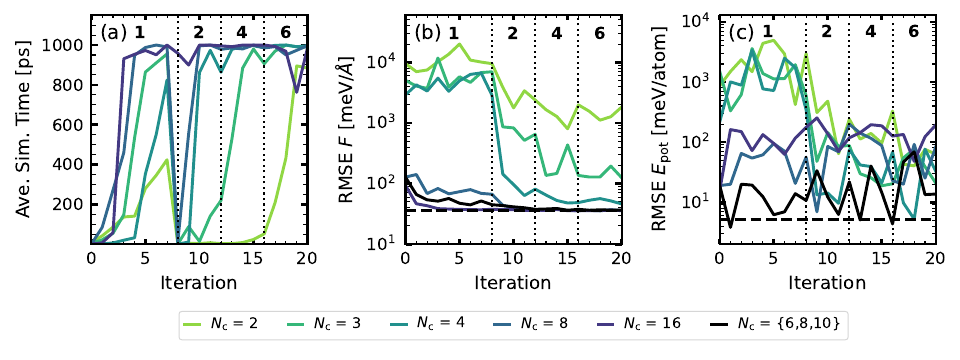}
		\caption{Evaluation of the active learning process of training ACE potentials on oligomers of a single specific length and a combination of 3 different length. a) Average simulation time of test simulations before hitting the stopping criterion due to extrapolation. The maximum simulation time is 1 ns. Evolution of b) the RMSE of forces and c) RMSE of atomic energy during the active learning process. The number of molecules employed in the test simulations is shown at the top of the figures. The dashed line represents the RMSEs of the refined ACE potential after iteration 20.}
		\label{fig:active_learning}
	\end{figure*}
	
	In the following, we evaluate the AL process as described above for training ACE potentials on oligomers of a specific chain length and evaluate the transferability of the potentials to oligomers of different sizes. Therefore, we define two different objectives which should be achieved during the AL process. 1) The potential must lead to stable trajectories for long timescales. 2) The accuracy must be sufficient. 
	
	We evaluate the first objective, by performing extended test MD simulations over 1 ns for each ACE potential trained during the AL loop. Note that these include explicitly not the refined potentials. The simulations were stopped if $\gamma_\mathrm{stop}>3.5$ was exceeded. Similar to the test simulations during the AL loop, they contain 1 molecule up to iteration 8, 2 molecules up to iteration 12, 4 molecules up to iteration 16 and 6 molecules beyond.
	
	\autoref{fig:active_learning}a depicts the average simulation over these simulations. For single molecule test simulations, the chain length has a small impact on the stability. 
	
	For most chain lengths $N_c=8$ and $N_c=16$ the target simulation time was almost achieved after five AL iterations. However, for smaller chain lengths the simulations remain partially unstable even after seven iterations. For $N_c<8$, there is no clear trend in the simulation stability, since the potentials for $N_c=3$ are more stable than the simulations for $N_c=4$. Only the simulations for $N_c=2$ are clearly the most unstable ones. 
	
	These differences may be attributed partly to statistical fluctuations, however there is also another possible explanation: While the number of initial training structures and structures added per AL iteration is identical, the chain length is different. Thus more training data is available in terms of local atomic environments with an increasing chain length.
	Contrary, larger chains have also more degrees of freedom (reflected by an increased coverage of the descriptor space) and are thus more difficult to characterize and require an increased amount of training data, such that there might be an optimal chain length in the sense that it balances the number of available local environments and degrees of freedom.
	
	When a second molecule is added for the test simulations but not yet for the training data (iteration 8) the average simulation time drops significantly. For chain lengths $N_c<16$ the simulations are very unstable with less than a nanosecond simulation time achieved on average. Only for $N_c=16$ there is just a small decrease in the stability, indicating that relevant intermolecular interactions have been learned already from a single chain. 
	
	This behavior is only possible for longer chains, which possess sufficient conformational flexibility to bend and allow chain ends or other segments to interact via non-bonded interactions. By adding a small number of structures containing clusters of two molecules to the training set (iteration 10), simulations for chain lengths $N_c=8$ and $N_c=4$ become stable. However, the chain lengths $N_c=3$ and $N_c=2$ require more structures to obtain stable simulations. This is attributed to a low density of smaller alkanes, which remain in the gaseous phase rather than forming clusters. Additionally to the smaller number of local atomic environments they provide just limited new information. To counteract this, increasing the number of molecules in the simulation box enhances the frequency of intermolecular interactions. Once they are adequately learned, including larger clusters in the training set does not further improve stability. Stability across all chain lengths is achieved after 20 AL iterations, except for $N_c=2$ where the average simulation time remains only close to 900 ns. Note that larger fluctuations for the stability, e.g. as observed for $N_c=16$ in iteration 19, may be caused by non-optimal fitted parameters of the ACE potential, which is conditional to the random initialization of parameters.
	
	The second objective is evaluated by calculating the RMSE of the forces and the potential energy on an external test set, which is similar to the set of structures employed in the PCA analysis. Because this test set of structures encompasses various chain lengths, the resulting RMSEs provide insight non only into the performance of the trained potential for a specific chain length but also into its transferability across different chain lengths.
	
	\autoref{fig:active_learning}b and c display the RMSE of the forces and the potential energies, respectively. Since $N_c=2$ occupies a different region in the descriptor space, RMSEs are expected to be higher if not trained specifically on this chain length. Therefore, the RMSEs were additionally calculated without considering $N_c=2$ in the test set, however, no significant impact on the RMSEs was found for $N_c\geq8$. Note that the RMSEs of the refined potential for $N_c=\{6,8,10\}$ after ifteration 20 is shown as a dashed line in these figures for reference.
	
	For the RMSE of the forces depicted in \autoref{fig:active_learning}b two distinct regimes are observed depending on the chain length. For the potential trained on $N_c<8$ the RMSE is larger than $2 \cdot 10^3\; \mathrm{meV\;\AA^{-1}}$, compared to $200 \; \mathrm{meV\;\AA^{-1}}$ for $N_c = 8$ and $100 \; \mathrm{meV\;\AA^{-1}}$ for $N_c=16$ in the initial iteration. 
	
	Note that RMSEs below $100 \; \mathrm{meV\;\AA^{-1}}$ are already within an acceptable range of errors, typically encountered in other MLIP applications \cite{Magdau.2023, Batatia.2025}. 
	
	Whereas for $N_c\geq 8$ a slight decrease of the RMSE during the initial 2 AL iteration can be observed, there is no such trend for smaller oligomers. While only a single molecule is present in the training set, the RMSE remains constant and only drops if clusters of two molecules are added to the training set. When comparing $N_c=2$, $N_c=3$ and $N_c=4$, the improvement increases for larger chain lengths and only for $N_c=4$ the RMSE is in an acceptable range below $100 \; \mathrm{meV\;\AA^{-1}}$ after 11 iterations. 
	For $N_c=8$ the RMSE decreases also further and converges to the RMSE observed for $N_c=16$, which is $36 \; \mathrm{meV\;\AA^{-1}}$. This is already close to the optimum since no significant further improvement is observed for these chain lengths during later AL iterations including also larger clusters in the training set (RMSE of $35 \; \mathrm{meV\;\AA^{-1}}$ after 20 AL iterations). For $N_c=4$ the RMSE improves further and equals $46 \; \mathrm{meV\;\AA^{-1}}$ after 20 AL iterations. Also for $N_c=3$ the RMSE of $126 \; \mathrm{meV\;\AA^{-1}}$ is still acceptable. Only for $N_c=2$ it is larger than $1800 \; \mathrm{meV\;\AA^{-1}}$ and therefore not applicable in any simulation. 
	
	Whereas it seems to be possible to train a potential for PE using only butane ($N_c=4$) in the training set, slightly larger oligomers require a significantly reduced number of training iterations. Based on these results either a single oligomer chain of hexadecane ($N_c=16$) or two oligomers of octane ($N_c=8$) seem to be optimal and reach a similar RMSE below $40 \; \mathrm{meV\;\AA^{-1}}$ after a few AL iterations. Potentially, the number of AL iterations can be further reduced for octane if a second molecule  is employed in the test simulations already during the first AL iterations. 
	
	For the RMSE of the energy, which is shown in \autoref{fig:active_learning}c, the trends are not such clear. This is caused  by a higher weighting of forces in the loss function, with the refined potential as an exception. However, for $N_c\geq8$ the RMSE is almost constant within fluctuations during all AL iterations, whereas it drops for $N_c<8$ after adding clusters of two oligomers to the training set. After 20 AL iterations the RMSE is nearly independent of the chain length employed in the training set. The observed RMSE fluctuates strongly between approximately $10\:\mathrm{meV/atom}$ and $200\:\mathrm{meV/atom}$. This is much higher than expected and found in literature for similar systems of small molecules. However, the high RMSE is mainly caused by a trivial shift in potential energy that depends on the ratio of carbon to hydrogen atoms and thus also on the chain length (see SI \autoref{si_fig:e_pot_ratio}) \cite{Hooven.2026}. 
	This shift cannot be captured by including only oligomers of a single chain length in the training set. However, by incorporating various chain lengths, the shift can be learned, and the RMSE is substantially reduced. We demonstrate this by training a potential on a combined set of $N_c=6,8\, \mathrm{and} 10$ (see refined potential), using an  AL scheme similar to that used for the individual chain lengths.
	
	Note that we have analyzed the choice of temperature in the test simulations as well as the choice of the cutoff radius for the ACE potential for collecting the data in a similar manner (see SI \autoref{si_fig:active_learning}). It turns out that including structures obtained from simulations with higher temperatures in the training set provides a generally better description for simulation at other temperatures than including only structures sampled at low temperatures. The cutoff radius has only minor impact. During the first AL iterations when only single molecules are sampled a smaller cutoff-radius leads to a faster decrease of the RMSEs. However, as soon as clusters of molecules are considered, a larger cutoff-radius becomes slightly beneficial.
	
	\subsection{Inter-molecular interactions}
	
	\begin{figure*}[h]
		\centering
		\includegraphics[width=0.5\linewidth]{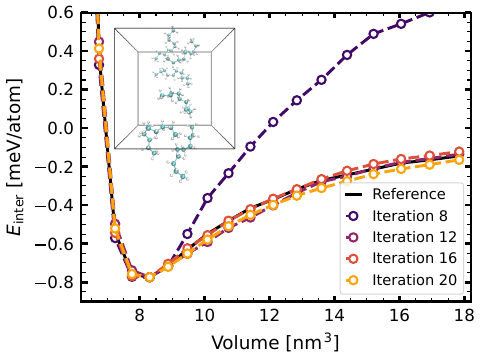}
		\caption{Inter-molecular potential energy for volume scans using refined ACE potentials from different iterations of the active learning loop. The reference is shown as a black line. All potentials are shifted such that their minima are aligned.}
		\label{fig:volume_curve}
	\end{figure*}
	
	In the following, we focus on the correct representation of intermolecular interactions, as they are critical to accurately describe the density of a bulk system; even slight deviations can have a significant impact on the computed density. 
	\autoref{fig:volume_curve} shows the intermolecular potential as a function of the simulation box volume for a cluster of six octane molecules. While scaling the volume of the box, all intramolecular distances were kept constant. The black line represents the ground truth reference intermolecular potential energy. The colored data points represent the energies obtained from ACE potentials at different stages of the active learning procedure, incorporating different chain length $N_c=6$, $N_c=8$ and $N_c=10$ in the training set. Since the potential energy offset can be chosen arbitrarily and does not affect the computed forces, all potentials were shifted to align their minima.
	
	While all ACE potentials accurately describe the interamolecular potential energy close to the equilibrium volume, significant deviations may emerge at larger volumes where intermolecular distances increase. Specifically, potentials fitted only on single molecules during early stages of the AL process (iteration 8) underestimate the intermolecular potential energy at larger volumes. This would likely result in an overestimated bulk density.
	When clusters of at least two molecules are included in the training set the predictions closely follow the ground truth reference, with some minor deviations emerging at increased volumes beyond the equilibrium, which is described by all potentials sufficiently well. There are no significant trends observed in the potentials fitted at later stages of the AL procedure. For example the potential fitted at iteration 12 is most accurate for the largest volumes, whereas the potential fitted at iteration 16 is most accurate for intermediate volumes. The potential fitted at iteration 20 is not optimal for both but provides a solid accuracy over the full volume range.
	Note that it is inherently important to refine the potential by shifting to a higher weighing of the energy in the loss function in a second step, since otherwise inaccurate predictions are obtained for large volumes where premature plateaus of the potential energy are observed (see SI \autoref{si_fig:volume_scan_si}).
	
	These results underscore the importance of an appropriate selection of training data and careful weighting of energies versus forces in the loss function. Furthermore, it highlights the necessity of of analyzing specific interaction components rather than relying solely on total energy and force errors, as deviations in intermolecular interactions can be masked by the dominating intramolecular part.
	
	Additional dissociation curves of two aligned octane molecules are shown in \autoref{si_fig:dissociation}. All trends remain similar, however, the predictions are less accurate and non-physical oscillations occur in the energy vs. volume curves caused by interpolation errors, since perfectly aligned alkanes are not part of the training set. This demonstrates also some limits of the refined ACE potential, which nevertheless shows reasonable interpolation capabilities for unseen structures. In amorphous structures the observed oscillations would be averaged out, even if they were still present.
	
	\begin{figure*}[h]
		\centering
		\includegraphics[width=\linewidth]{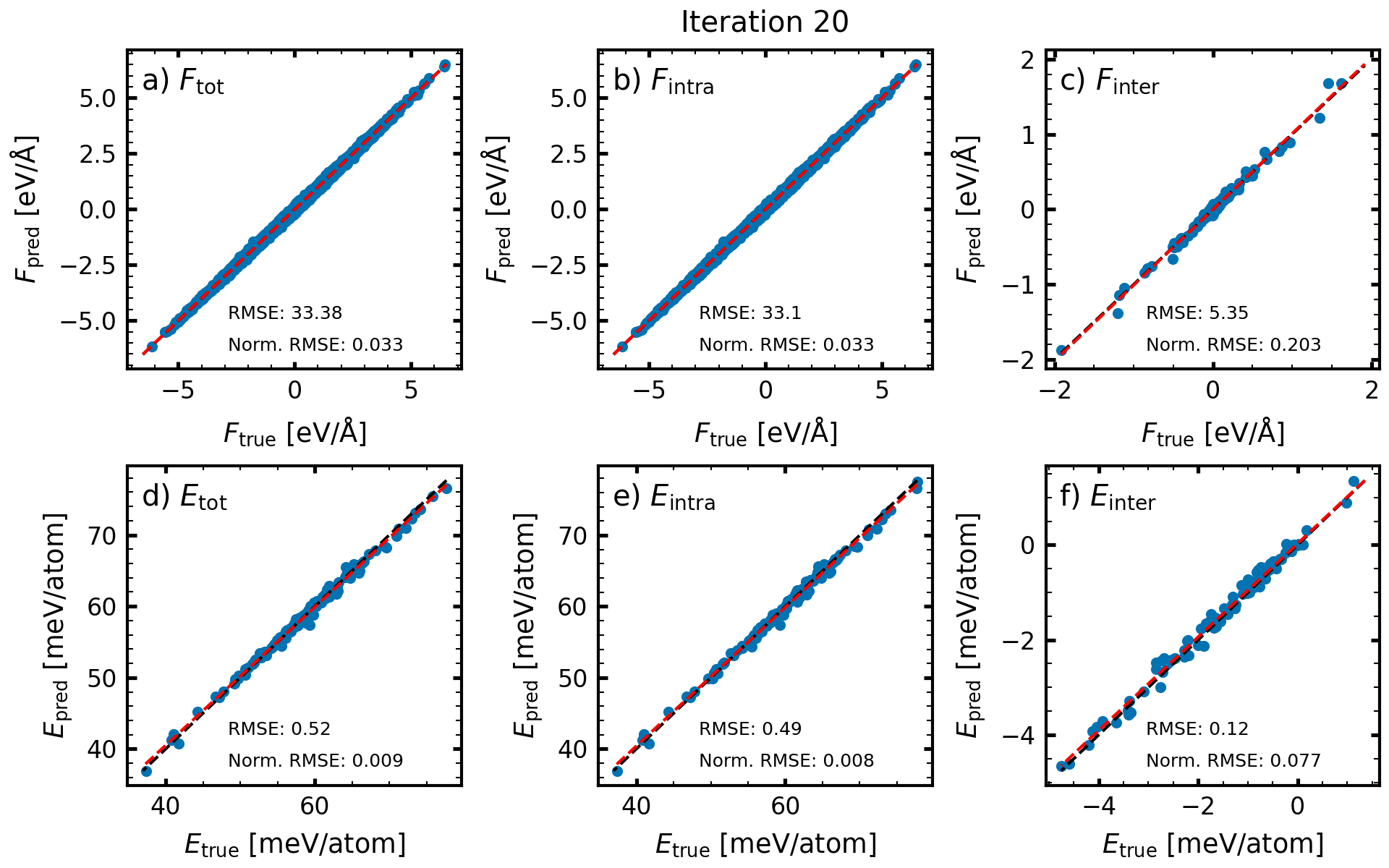}
		\caption{Pairplots of predicted versus true a) total forces, b) intramolecular forces, c) intermolecular forces, d) total energies, e) intramolecular energies and f) intermolecular energies for the final ACE potential. RMSE and normalized RMSEs are provided for each contribution. The normalized RMSEs are calculated by dividing the RMSE by the root mean squared reference forces and energies. The black dashed lines represent a perfect fit, the red dashed lines repersent a linear fit of predicted \textit{vs.} true forces and energies.}
		\label{fig:pairplots}
	\end{figure*}
	
	While a single volume scan is illustrative to understand the fundamental behaviour of fitted potentials, it is not sufficient to evaluate the accuracy and compare different potentials. Thus, we performed a more detailed analysis of the predictive performance for inter- and intramolecular interactions by calculating the individual contributions for the test set of the refined potential after iteration 20, which comprises 5 \% of all structures collected during the AL procedure.
	
	\autoref{fig:pairplots}a-c shows the predicted total, intra- and intermolecular forces and energies plotted vs. the ground truth reference for the refined potential at this iteration. Generally, a good agreement between predictions and ground truth reference is observed. The agreement is in particular excellent for the intramolecular forces (R2-score of 0.999), while the intermolecular forces have a slightly lower R2-score of 0.958.
	
	While the R2-score is an intuitive first metric, it lacks the capability of quantifying the exact error of the potentials. Complementary to the R2-score, the root mean squared error (RMSE) of the forces and energies provides exactly this information. However, since intramolecular forces and energies are typically orders of magnitudes larger than the intermolecular contributions, the RMSEs for both are not directly comparable. Therefore, we introduce a normalized RMSE, which is calculated by dividing the RMSE by the reference root mean squared energies or forces, respectively. 
	
	When evaluating the RMSEs for different AL iterations (see \autoref{tab:table_metrics} and SI \autoref{si_sec:pairplots} for the pairplots), only a very small improvement is observed for intramolecular forces from 36.9 $\mathrm{meV\;\\A^{-1}}$ after iteration 8 to 33.1 $\mathrm{meV\;\\A^{-1}}$ after iteration 20. For the intermolecular interactions the RMSE decreases from 21.3 $\mathrm{meV\;\\A^{-1}}$ after iteration 8 to 7.3 $\mathrm{meV\;\\A^{-1}}$ after iteration 12 and then only decreases slowly to 5.4 after iteration 20, indicating that the main improvement emerges when adding clusters of two atoms to the training set. Also for the total forces, the main improvement is observed when going from iteration 8 to iteration 12. The trends are also resembled in the normalized RMSE, allowing to compare the relative error of intra- and intermolecular forces. For the intramolecular forces a normalized RMSE between 0.037 and 0.033 is observed, which is much smaller than the normalized RMSE for intermolecular forces with 0.808 after iteration 8 and 0.203 after iteration 20. This indicates that significant deviations in the predictions of the intermolecular forces are still present. They could be attributed to the fact that the ACE potential applies a cutoff radius such that interactions beyond can be only considered in a mean field manner. This gives rise to a strongly increased normalized RMSE, compared to intramolecular interactions, which are all captured within the cutoff radius.
	
	The energetic contributions corresponding to the forces are also shown in \autoref{fig:pairplots}d-f.
	Whereas for the refined potentials after iteration 16 and 20 good agreement between predicted and true intra- and intermolecular interactions is observed, similar pairplots for refined potentials fitted at earlier stages of the AL process directly show larger deviations especially for the predicted intermolecular energy compared to the ground truth reference (see SI \autoref{si_sec:pairplots}). In particular, the slope of predicted vs. true intermolecular energies is significantly larger than one after iteration 8 and significantly smaller after iteration 12. This is also directly reflected by a significantly higher RMSE of the intermolecular energy, compared to later iterations. 
	
	When considering only the energetic contributions, the RMSEs of all interaction types are smaller after iteration 16 than after iteration 20, e.g. the RMSE of the intermolecular energy is 0.09 $\mathrm{meV\:atom^{-1}}$ and 0.12 $\mathrm{meV\:atom^{-1}}$, respectively. This could be explained by the fact that primary forces where optimized during the first step of the potential training and furthermore different local minima of the ACE parameters are discovered, leading to slightly different RMSEs close to the optimum RMSE, when training with different training data. 
	
	Overall, the refined potential after iteration 20 reproduces intra- as well as intermolecular interactions with high accuracy. Since there is only a small improvement in the force predictions and no improvement in the energy predictions when adding more data from iteration 16 to iteration 20, improving the ACE potential further would require an disproportional large additional amount of data. Moreover, the potential is mostly not limited by data but by the choice of the complexity of the potential and the cutoff radius itself. 
	
	\begin{table}[]
		\centering
		\caption{Comparison of the RMSE and RMSE normalized by the root mean squared forces and energy, respectively, for total, intra- and intermolecular forces and energies. Refined potentials were fitted to datasets obtained after different iterations of the AL process. The number of local atomic environments in each training data set is also provided. 
			The RMSEs were evaluated for the same set of structures (test set for the refined potential fitted after iteration 20). }
		\begin{small}
			\begin{tabular}{l l | c c c | c c c | c c c | c c c}
				\hline
				Iteration & \multicolumn{1}{p{1.75cm}|}{Atomic \:\:\:\:  environments} &
				\multicolumn{3}{p{2.7cm}|}{\centering RMSE Forces [$\mathrm{meV\;\AA^{-1}}$]} &
				\multicolumn{3}{p{2.7cm}|}{\centering RMSE Energy [$\mathrm{meV\;atom^{-1}}$]} &
				\multicolumn{3}{p{2.7cm}|}{\centering Normalized. RMSE Forces} & 
				\multicolumn{3}{p{2.7cm}}{\centering Normalized. RMSE Energy} \\
				& & Total & Intra & Inter & Total & Intra & Inter & Total & Intra & Inter & Total & Intra & Inter \\
				\hline
				8 & 21644    & 42.6 & 36.9 & 21.3 & 1.4 & 0.44 & 1.3 &
				0.042 & 0.037 & 0.808 & 0.023 & 0.007 & 0.849 \\
				12 & 58230   & 34.8 & 34.2 & 7.3 & 0.62 & 0.51 & 0.30 &
				0.034 & 0.034 & 0.278 & 0.010 & 0.008 & 0.192 \\
				16 & 139584  & 34.5 & 34.2 & 5.7 & 0.50 & 0.48 & 0.09 &
				0.034 & 0.034 & 0.215 & 0.008 & 0.008 & 0.057 \\
				20 & 246238  & 33.4 & 33.1 & 5.4 & 0.52 & 0.49 & 0.12 &
				0.033 & 0.033 & 0.203 & 0.009 & 0.008 & 0.077 \\
				\hline
			\end{tabular}
		\end{small}
		\label{tab:table_metrics}
	\end{table}

	\subsection{Density}
	\begin{figure*}[h]
		\centering
		\includegraphics[width=1.0\linewidth]{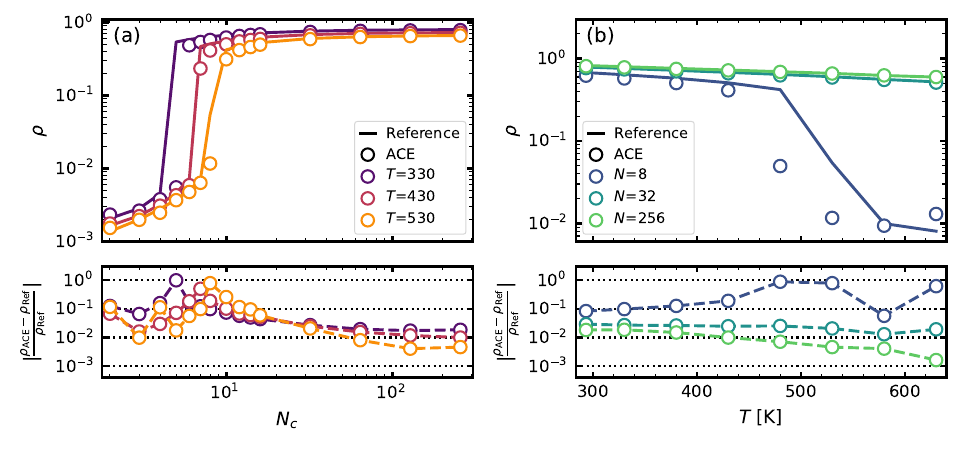}
		\caption{Comparison of densities in various bulk simulations with the reference and the ACE potential. a) Dependence of the density $\rho$ on the chain length for fixed temperatures and b) dependence of $\rho$ on the temperature for fixed chain lengths. The lower plots show the relative difference in densities between systems simulated with the ACE and the reference potential.}
		\label{fig:density}
	\end{figure*}
	
	The density can be considered as a central property which needs to be reproduced correctly as a prerequisite to reproduce also other structural and dynamical properties correctly. The accuracy of the refined ACE potential after iteration 20 is high enough to expect that also the density as a bulk property, which is very sensitive to intermolecular interactions, should be reproduced sufficiently well. However, the exact effect of the error in the predictions of intermolecular interactions remains unclear so far. 
	
	To check this, we performed a series of bulk simulations with the reference and with the refined ACE potential after iteration 20 and plot the observed density \textit{vs.} chain length for fixed temperatures and density \textit{vs.} temperature for fixed chain length in \autoref{fig:density}a and b, respectively. Exact values of the densities are provided in \autoref{fig:density_si}.
	
	While smaller oligomers (up to $N_c=4$ at 300 K and $N_c=8$ at 530 K) exist in the gaseous phase, larger oligomers and polymers are in the liquid phase (\autoref{fig:density}a). Although the potential was trained on small oligomers between $N_c=6$ and $N_c=10$, which includes both, gaseous and liquid phase structures around $T_\mathrm{train}=430\;\mathrm{K}$, it yields more accurate densities for larger polymers. Depending on the temperature, the relative difference between the reference and ACE densities ranges from 1 \% and 10 \% for small oligomers, peaks at nearly 100 \% where the gas-to-liquid transition occurs and subsequently declines to below 2 \% for $N_c>100$. At $T=530\;\mathrm{K}$ the deviation is further reduced to below 0.5 \%. 
	A more detailed evaluation of the temperature dependence of the density is represented in \autoref{fig:density}b, showing that the ACE potential reproduces the reference mostly well, particularly for larger polymers. For $N_c=256$, the relative difference is 3 \% at $T=293\;\mathrm{K}$ and drops below 2 \% for $T>430\;\mathrm{K}$.
	For intermediate chain lengths, such as $N_c=32$, the deviations remain almost constant close to 2-3 \% regardless of the temperature. 
	Smaller oligomers ($N_c=8$) exhibit significantly larger errors, starting at 10 \% at $T=293\;\mathrm{K}$ and peaking at nearly 100 \% close to the phase transition temperature. In the gasous phase the density difference fluctuates between 10 \% and 100 \%.
	
	Although we are not primarily interested in simulating small oligomers in this study, these observations clearly show some limitations of the fitted potential. The comparatively lower density close to phase transition temperatures in simulations with the ACE potential are mainly caused by neglected long-range interactions due to the introduction of a cutoff radius of $7\;\mathrm{\AA}$. The mean-field treatment of these long-range interactions within the ACE potential breaks down when the distances between molecules increase and also small contributions become important.
	Similar observations were already made when decreasing the cutoff-radius in simulations with the classical OPLS-AA force field \cite{}.
	
	In contrast, if the molecules are sufficiently large such that the intermolecular interactions are strong enough to keep molecules close to each other even at elevated temperatures, the deviations in the predicted intermolecular interactions become less important. Furthermore, for larger molecules also topological restrains play a more important role, which keep molecules also close to each other due to entanglements. 
	As a summary, even minor inaccuracies in the prediction of intermolecular interactions can lead to fundamentally different densities for small molecules, whereas larger molecules are less sensitive to such deviations. 
	
	\subsection{Structure}
	\begin{figure*}[h!]
		\centering
		\includegraphics[]{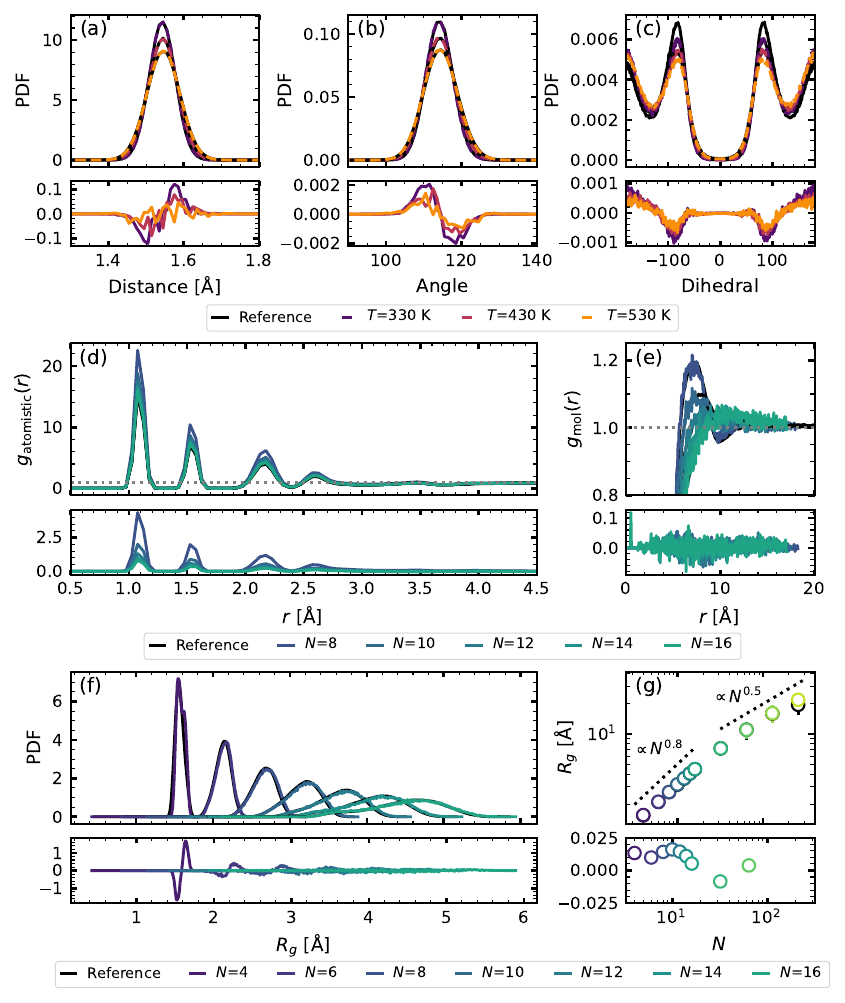}
		\caption{Comparison of structural properties between systems simulated with the reference OPLS-AA and the fitted ACE potential. Distribution of carbon backbone a) bond distances, b) angles and c) dihedrals in octane. d) atomistic and e) molecular radial distribution functions of selected oligomers at 430 K. f) Distribution of the radius of gyration $R_g$ for oligomers of different length. e) Dependence of the $R_g$ on the chain length of the polymer at 430 K. The lower panel of each subplots shows the difference between the ACE and reference potential.}
		\label{fig:structure}
	\end{figure*}
	
	We evaluated several structural properties of the oligomer and polymer bulk systems, including bond, angle and dihedral distributions of the polymer backbone to characterize the structure of individual chain segments, the full atomistic and molecular radial distribution functions $g_\mathrm{atomistic}(r)$ and $g_\mathrm{mol}(r)$, respectively, as well as the radius of gyration to characterize the chain structures in bulk. The key results are displayed in \autoref{fig:structure}, where the upper panel of each subplot shows the specific structural property obtained with the ACE and the reference potential and the lower panel shows the difference between the ACE and the reference potential.
	
	The local structure of individual chain segments is well reproduced by the ACE potential as depicted in \autoref{fig:structure}a-c for the bond distance, the angles and dihedrals for the carbon backbone atoms. The distributions obtained from the ACE and reference potential are almost identical for a broad temperature range, exemplary shown for temperatures between 330 K and 530 K for $N_c=8$, although the ACE potential was only trained on structures generated at 430 K. 
	For even lower temperatures, it is expected that the distributions are still well represented by the ACE potential, however, for higher temperatures the potential is expected to extrapolate significantly, since strong non-equilibrium structures are encountered, which are not well represented in the training set. This can be seen in reduced simulation times due to high extrapolation grades for $T>530\;\mathrm{K}$. For these simulations, statistics are not reliable enough to obtain meaningful distributions (see SI \autoref{fig:sim_time}). 
	When evaluating the difference in the distributions between the ACE and reference potential more carefully (lower panels of \autoref{fig:structure}a-c) it becomes evident that a small asymmetry is present in the ACE potential. For example, close to the equilibrium bond distance ($1.54\;\mathrm{\AA}$), larger bond distances are slightly over-represented and smaller bond distances are slightly under-represented in the ACE potential, caused by an underlying small shift of the potential energy distribution (see SI \autoref{si_fig:dissociation}). A similar asymmetry appears in the angular distribution (equilibrium angle of $114\;^\circ$), with larger deviations than those seen in bond distances. While those deviations are negligible, the dihedral distributions show a notable discrepancy, with a significantly reduced probability at the equilibrium dihedral ($\pm 80\;^\circ$) compared to the reference. 
	
	Similar observations were made for angles, bonds and dihedrals including hydrogen atoms, with differences in the distributions being notably smaller. These differences are nearly indpendent of the chain length (see also \autoref{fig:structure_temp} and \autoref{fig:structure_Ns} in the SI).
	
	The atomistic radial distribution function $g_\mathrm{atomistic}(r)$ between central carbon atoms and other carbon and hydrogen atoms is depicted in \autoref{fig:structure}d for various chain lengths in liquid bulk systems. The first peaks are attributed to intra-chain interactions, for example, the first peak is attributed to the carbon-hydrogen bond and the second peak to the carbon-carbon bond. The third peak refers to the interaction between a central carbon atom and a hydrogen atom bonded to a neighboured carbon atom. 
	Since $g(r)$ is normalized by the system density, the peak intensity varies, decreasing as the density - and thus the chain length - increases. Overall, the ACE potential shows good agreement with the reference. However, the peaks in the ACE potential are slightly overestimated. This discrepancy diminishes with increasing chain length, which can be attributed to the improved accuracy of densities for longer chains.
	The molecular radial distribution function $g_\mathrm{mol}(r)$ between the center of masses of a central molecule and of surrounding molecules is depicted in \autoref{fig:structure}e. 
	Whereas for smaller chain lengths ($N_c\leq 12$) a distinct small peak around $7\mathrm{\AA}$ to $9\mathrm{\AA}$ is observed, it is shifted to higher distances and diminishes nearly as the chain length increases. This is caused by less ordered structures and an increasing number of entanglements (see below).
	The number of molecules in a simulation box decreases with increasing chain length, therefore less data is present and $g_\mathrm{mol}(r)$ is noisy for the longest chains. This makes it impossible to compare the ACE and reference potentials with respect to small differences. However, similar to $g_\mathrm{atomistic}(r)$ there is generally good agreement with the reference also for $g_\mathrm{mol}(r)$.
	
	Finally, the radius of gyration $R_g$ is evaluated. \autoref{fig:structure}f and g display the distribution of $R_g$ for various chain lengths in gas and liquid phases, as well as the dependence of $R_g$ on the chain length, respectively.
	Overall, the distributions for ACE and the reference show good agreement. However, the ACE potential exhibits a small shift toward slightly larger $R_g$, which is consitent with the previously observed overestimation of carbon-carbon bond distances.
	$R_g$ increases with an increasing chain length, following a power law $R_g\propto N_c^\beta$, where two regions with different exponents $\beta$ can be identified. Up to $N_c\leq16$ alpha is approximately 0.8, and for $N_c\geq32$, $\beta=0.5$ is observed, as would be expected for polymers \cite{Flory.1954}.
	Whereas for small oligomers $R_g$ is slightly overestimated, $R_g$ agrees better in the second regime. However, larger deviations are observed for $N_c=512$. This can be attributed to insufficient averaging and finite-size effects which are generally more pronounced in systems with longer chains.
	
	\subsection{Dynamics}
	
	\begin{figure*}[h]
		\centering
		\includegraphics[width=\linewidth]{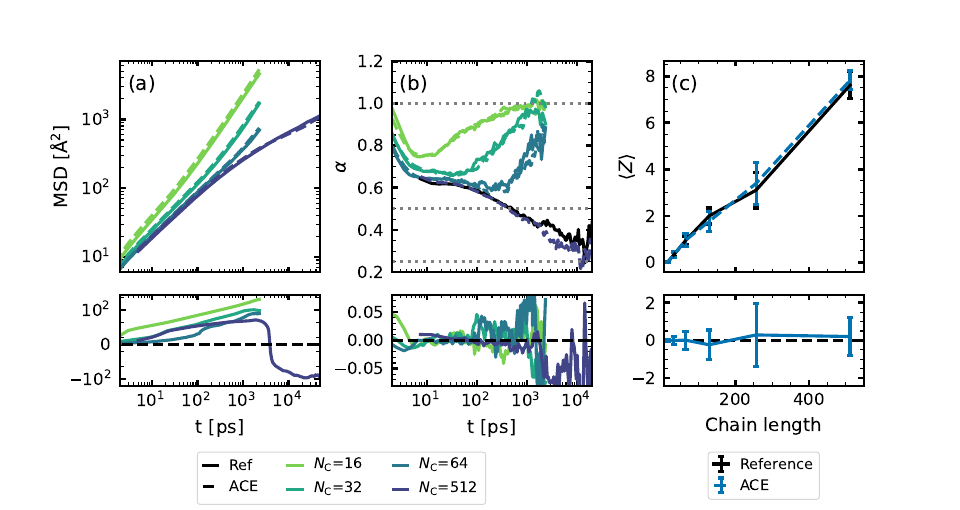}
		\caption{a) Mean squared displacement of the central carbon atoms of the polymer for different chain length at $T=430\,\mathrm{K}$. Those 50 \% of the carbon atoms were selected which are most central within each chain. b) Tracking of $\alpha$ to discriminate different regimes of subdiffusive behaviour. c) Dependence of the average number of entanglements $\langle Z \rangle$ on the chain length.}
		\label{fig:msd}
	\end{figure*}

	Having characterized the accuracy of the structures reproduced by the ACE potential, we now investigate the dynamics with a focus on the segmental mean squared displacement (MSD). \autoref{fig:msd}a displays the MSD for the carbon backbone atoms of four different chain lengths. To minimize chain-end-effects but still ensure statistical reliability, only the central 50 \% of the carbon atoms in each chain were included in the analysis. 
	
	All displayed bulk systems are in an amorphous or liquid state. Generally, the MSD increases faster for larger chain lengths. This trend and the overall MSD is accurately reproduced by the ACE potential. However, similar to the structural properties, also for the dynamics small deviation occur. While the MSD obtained from simulations with the ACE potential is larger for all chain length up to $N_c=256$, it is smaller for $N_c=512$ after 10 ns. The deviations might be attributed partly to statistical fluctuations; however, they can also be explained by slightly lower densities in simulations with the ACE potential, which leads to an increased dynamics.
	
	To understand the different regimes of the MSD and their proportionality ($\mathrm{MSD}\propto t^\alpha$) it is helpful to characterize the entanglements of the oligomers and especially polymers, which strongly influence the dynamics. 
	
	Compared to small oligomers, the dynamics of individual central polymer segments are restricted by their covalent connectivity to adjecent segments. This is typically evident in the MSD, where a sub-diffusive regime with $\alpha=0.5$ emerges. Entanglements further constrain the dynamics, as the segments are restricted to move along tubes formed by surrounding entangled polymers. This results in a second, slower sub-diffusive regime with $\alpha=0.25$. Consequently, nanoseconds long simulation times are essential to fully characterize these dynamical regimes. \cite{Cates.1987, Kremer.1990}
	Note that the mentioned exponents $\alpha$ may be only obtained for sufficiently long chains. Segments at the chain ends are less restricted in their movement and therefore higher values of $\alpha$ are expected.
	
	Generally, it is important that the entanglements are also reproduced accurately by the ACE potential. This requires that all bonds are maintained during the simulation and no bond-breaking or new bond formation occurs, which may lead to a different entanglement situation. Entanglements typically occurs for polymers larger than their corresponding, polymer-specific entanglement length $N_e$. To observe the sub-diffusive regime of $\alpha=0.25$ in the MSE, typically chain lengths of approximately $5\; N_e$ to $10\; N_e$ are necessary \cite{Behbahani.2025}.
	
	As shown in \autoref{fig:msd}c, the average number of entanglements per chain $\langle Z \rangle$ depends approximately linear on the chain length. Within error bars, the ACE and the reference potential give the same values for $\langle Z \rangle$, indicating a well-preserved topology in the ACE potential, as already discussed in more detail when comparing other structural properties.
	
	From this linear dependence of $\langle Z \rangle$, $N_e$ can be estimated \cite{Hoy.2009}. Here, we obtain $N_e \approx 60$, which is in agreement with the literature values for polyethylene \cite{Foteinopoulou.2006, Hoy.2009, Litvinov.2013}, although densities are different due to missing long-range vdw-interactions in this work. Consequently $N_c > 300$ is necessary to observe the relevant $\alpha$-regime.
	
	Due to long simulations times that might be necessary to observe this regime and high computational costs of the ACE potential, we decided to run an extended simulation for more than 100 ns only for $N_c=512$ to demonstrate that our fitted ACE potential reproduces the dynamics of entangled polymers correctly. 
	
	When evaluating the dependence of $\alpha$ on lag time $t$, which can be calculated from the MSD, the different regimes become visible as plateaus (see \autoref{fig:msd}b). 
	Independent of the chain length the first plateau is observed after approximately 10 ps at $\alpha\approx0.75$ for $N_c=16$ and decreases with increasing chain length to $\alpha\approx0.6$ for $N_c=512$. The deviations from $\alpha=0.5$ can be explained partly by small chain length and partly by the dependence of the segment dynamics on the position of the segment within a chain. For finite chain length, $\alpha=0.5$ would be observed only for the most central segments. Since we average over the 50 \% of most central carbon atoms, $\alpha$ is increased compared to averaging over a smaller portion carbon atoms. However, the statistical description is not sufficient due to a smaller amount of available data.
	
	For smaller chains with $N_c \leq 64$ $\alpha$ increases after the first plateau and approaches the diffusive regime ($\alpha=1$). Only for the largest chain length $N_c=512$ $\alpha$ decreases further and reaches a second plateau after 10 ns at $\alpha=0.3$, indicating that a restricted dynamics due to entanglements is present. Note that longer simulation times would be needed to increase the statistical description and decrease the noise observed for these time scales. 
	
	Similar values of $\alpha$ are obtained with the ACE and the reference potential. Especially for the first plateau an excellent agreement is observed. For the second plateau for $N_c=512$ a reliable comparison is not possible due to large fluctuations in $\alpha$. Overall the ACE potential reproduces the dynamics of the reference systems extremely well.
	
	\section{Conclusions}
	
	We have presented a systematic study on MLIPs for polymer systems and identified ACE as an ideal descriptor. Already very small systems like clusters of only two small oligomers (e.g. $N_c=8$) or one larger oligomer ($N_c=16$) are sufficient to cover the full relevant descriptor space encountered when simulating various different systems, including larger chains or bulk systems with different densities.
	We implemented an automated active learning scheme, utilizing ACE and Lammps to efficiently create training data in an iterative way. Already after a few iterations, stable ACE potentials were obtained for single molecule systems. By including clusters of two or more molecules simulations were also stable for much larger clusters.
	Whereas the data collected during the later active learning iterations does not improve the predicted total forces and energies significantly, it still improves the inter-molecular energies and forces. By refining the potential after the final iteration the prediction accuracy can be improved further. 
	Whereas deviations of the predictions from the ground truth might not be clearly pronounced in the inter-molecular forces, they are directly visible in the intermolecular energy. During the AL iterations, the R2-score increases from 0.908 after iteration 12 to 0.958 for the refined potential when considering the intermolecular forces but from 0.583 to 0.991 for the intermolecular energy.   
	Although ultimately only forces are relevant in the MD simulations, considering metrics related to the potential energy can therefore provide further important insights.
	
	We demonstrated that it is crucial to predict correct intermolecular interactions to also reproduce all other properties observed in bulk simulations correctly. Even small deviations of the predicted intermolecular interactions from the ground truth reference can lead to significant changes in the density.  
	In particular, for small molecules the changes in the density can be drastic, being most pronounced close to the phase transition temperature between liquid and gaseous phase. Larger molecules are less affected and thus the fitted ACE potential provides more reliable results. 
	
	If energies and forces are fitted with high accuracy, all other properties are  reproduced in simulations with the ACE potential with high accuracy, too, compared to the reference potential. When evaluating structural properties, like bond-distance, angle and dihedral distributions as well as the number of entanglements, it is evident that the refined ACE potential maintains the polymer topology during the full simulation time. Furthermore, the dynamics is reproduced correctly, independent of the chain length and the corresponding number of entanglements per chain. 
	
	Overall, this study demonstrates the possibility to develop MLIPs for polymers based on clusters of small oligomers, which significantly reduces the computational costs in comparison to full bulk systems as training data. Therefore, we believe that our systematic study provides meaningful insights to enhance the development of MLIPs also for other polymeric systems and facilitate the efficient selection of training data to create large polymer databases. These might be used in future to create specialized or foundation models for polymers. 
	In this sense, our work paves the way to train MLIPs with near-ab initio accuracy for polymeric systems. A logical progression of this work would be to replace the force-field-based reference calculations with high-fidelity density functional theory calculations within our learning scheme for ACE potentials.
	
	\section*{Computational Details}
	
	\subsection*{Principal Component Analysis}
	For all structures within the given dataset (see \autoref{si_sec:dataset_descriptor}) the local atomic descriptors were calculated first using scikit-learn \cite{Pedregosa.2011}. The PCA was then performed using all local atomic environments to identify the two most important principal components.
	
	For each value of $N_c$ or $N_\mathrm{mol}$ clusters of individual atom types were identified using the DBSCAN \cite{Ester.1996} cluster algorithm to remove potential outliers. Afterwards the concave hull was calculated using the python library alphashape \cite{Bellock.2021}. 
	More details can be found in the example notebook.
	
	\subsection*{Local atomic descriptors}
	
	\subsubsection*{Smooth Overlap of Atomic Positions}
	
	The gaussian smoothed atomic density of species $Z$ is defined as
	\begin{equation}
		\rho^{Z_i}(r) = \sum_{j} \exp{\left(-\frac{|r-r_{ij}|^2}{2\sigma^2}\right)}
	\end{equation}
	which ensures invariance to permutation of equivalent atoms.
	It can be expanded with spherical harmonics $Y_{lm}(\theta,\phi)$ in an orthogonal basis
	\begin{equation}
		\rho^{Z_i} = \sum_{nlm} c^{Z_i}_{nlm} Y_{lm}(\theta,\phi)g_n(r)
	\end{equation}
	where $g_n(r)$ are radial basis functions, which are chosen as spherical gaussian type orbitals here. Given $c_{nlm}$ the power spectrum 
	\begin{equation}
		p_{nn'l}^{Z_i, Z_j} = \pi\sqrt{\frac{8}{2l+1}} \sum_{m} c^{Z_i}_{nlm} c^{Z_j}_{nlm}
	\end{equation}
	can be calculated which serves as the Smooth Overlap of Atomic Positions (SOAP) descriptor. \cite{Bartok.2013, Darby.2022}
	We compute SOAP descriptors using the DScribe library \cite{Himanen.2020, Laakso.2023}, employing a cutoff radius for local densities of $7\:\mathrm{\AA}$ and values of $n_\mathrm{max}=4$  and $l_\mathrm{max}=3$.
	
	\subsubsection*{Local Many Body Tensor Representation}
	
	Within this paper, we use a Local Many Body Tensor Representation (LMBTR) of second order, which is given by: 
	\begin{equation}
		L_2^{Z_i, Z_j}(r) = \sum_{j} w(r_ij)\exp{\left( - \frac{r-r_{ij}^{-1}}{2\sigma^2} \right)}
	\end{equation}
	with a weighting function 
	\begin{equation}
		w(r) = e^{-\frac{1}{2} r}.
	\end{equation}
	$L_2^{Z_i, Z_j}(r)$ was sampled at 100 discrete points of $r$ between $1/r_\mathrm{cut} < r < 1.05$ using the DScribe library implementation. Consistent with the other descriptors we chose $r_\mathrm{cut}=7\:\mathrm{\AA}$.
	
	\subsection*{Molecular Dynamics Simulations}
	All molecular dynamics simulations were performed using the LAMMPS software \cite{Thompson.2022} and the OPLS-AA force field parameters \cite{Jorgensen.1996} for reference simulations and fitted ACE potentials for all other simulations. 
	The ACE potentials were evaluated using the PACE package \cite{Lysogorskiy.2021} of LAMMPS.
	
	Different simulation setups were utilized for test simulations during the AL procedure and for subsequent simulations with the ACE potential and the OPLS-AA force field.
	
	\subsubsection*{Test simulations during AL}
	During the AL procedure, an equilibration with a timestep of 0.5 fs was performed for 500 ps with the OPLS-AA parameters, followed by a 50 ps production run with the fitted ACE potential in the NVT ensemble.
	
	The temperature for both, equilibration and production run, was varied within different active learning setups and managed by a Nose-Hoover thermostat \cite{Nose.1984}. A cubic simulation box with a size of $50\,\AA$ in any dimension was employed for single molecules to ensure that they effectively were treated as if they were in vacuum. For clusters of larger molecules, a box length of $30\,\AA$ was selected such that nearly isolated clusters and low density bulk structures are simulated.
	All structures with an extrapolation grade $\gamma_\mathrm{max}>2.0$ have been dumped for later consideration as additional training data and simulations have been interrupted if $\gamma_\mathrm{max}$ exceeds a threshold of 3.5.
	
	\subsubsection*{Simulations for system analysis}
	The following protocol was applied to all bulk simulations of polymers up to \ce{C256H514} analyzed in this work. A detailed setup of the analyzed systems is provided in \autoref{tab:sim_setup}.
	
	First, the systems were equilibrated in the NPT ensemble at an elevated temperature of 630 K and a pressure of 1000 bar for 500 ps. Temperature and pressure were controlled by a Nose-Hoover thermostat and a Parrinello-Rahman barostat \cite{Parrinello.1981}, respectively, utilizing the equations of motion by Shinoda \cite{Shinoda.2004} as implemented in LAMMPS. Coupling constants of 0.5 ps for temperature and 2.0 ps for pressure were employed.
	
	In a second step the system were cooled over a period of 100 ps to the target temperature, while the pressure was simultaneously set to 1 bar. All systems were then equilibrated for an additional 100 ps at the target temperature and ambient pressure of 1 bar. All equilibration steps were performed using the OPLS-AA force field. As target temperatures 293.15 K and 330 K to 630 K in steps of 50 K were selected.
	
	The production runs were initiated from these equilibrated structures, employing either the OPLS-AA force field or the refined ACE potential within the same NPT ensemble as during the last equilibration step. Each production run was simulated for 10 ns.
	
	For all stages, a time step of 1 fs was used. For OPLS-AA simulations, a cutoff-radius of $10\,\mathrm{\AA}$ was applied to van-der-Waals and electrostatic interactions. Long-ranged electrostatic interactions beyond were handled by the particle-particle particle-mesh (pppm) solver \cite{Hockney.2021}. 
	
	For simulations with the refined ACE potential, the cutoff-radius was determined by the potential itself ($7\:\mathrm{\AA}$). Due to the inherent locality of ACE potentials, long-ranged interactions were not considered. If $\gamma_\mathrm{max}$ exceeded a threshold of 25, the simulation was terminated, and the trajectory was analyzed up to that point.
	
	\subsubsection*{Extended simulations for the largest system}
	Unlike small oligomers, polymers require significantly longer equilibration and production time scales to accurately capture their equilibrium structure and dynamics. Consequently, extended simulations were performed for the longest polymer chain (\ce{C512H1026}), consisting of 25 polymer chains placed in a cubic simulation box with an initial side length of $150\:\mathrm{\AA}$. 
	
	The system was first relaxed via an VT simulation at 630 K for 2 ns. 
	The simulation box was then compressed in two stages using the NPT ensemble: First at 1000 bar for 10 ns, followed by 10 ns at 10 bar. 
	Subsequently, the structures were equilibrated at 630 K for 100 ns and then cooled to 430 K over a period of 10 ns. Finally, production runs were performed in the NPT ensemble for 100 ns using both, the OPLS-AA force field and the refined ACE potential. All other simulation parameters remain consistent with those described previously.
	
	\subsubsection*{Note on reference simulations}
	Typically, DFT or wavefunction-based methods are employed to calculate reference energies and forces for all structures in the training set \cite{}. However, the high accuracy of these methods is associated with significant computational costs, rendering nanosecond-scale simulations of large systems containing thousands of atoms computationally prohibitive.
	
	Polymer systems typically encompass significantly more degrees of freedom than small molecular systems and thus require extended simulation timescales to characterize their dynamics, which are governed by slow relaxation processes. Consequently, it is impractical to directly compare the dynamics of an MLIP-driven simulation with a corresponding high-accuracy reference simulation over these timescales.
	
	Nevertheless, when evaluating the transferability of potentials - such as testing a potential fitted to a specific oligomer length and temperature on systems of different lengths or temperatures - the absolute accuracy of the reference method is less critical than the consistency of the underlying physics. We expect that transferability is largely independent of the reference method's absolute accuracy, provided that essential interactions are adequately captured.
	
	Accordingly, this study focuses on a representative, yet simple  model system and utilizes the OPLS-AA force field as a computationally efficient reference. This approach enables nanosecond-scale reference simulations of large-scale systems, allowing for a rigorous evaluation of the MLIP's performance with regards to structure and dynamics. 
	
	\subsection*{Calculation of entanglements}
	The number of entanglements per chain was calculated by computing primitive paths and subsequently counting all kinks per chain as introduced by \cite{Kröger.2005} and implemented in the Z1+ software \cite{Kröger.2023}. 
	
	Primitive paths are represent the shortest paths that connect chain or segment ends without without inter-penetration of other principle paths \cite{Shanbhag.2007, Sukumaran.2005, Everaers.2004}. Kinks occur when chains cross each other and therefore are a good estimation to characterize the entanglements.
	
	Following the approach by Hoy \textit{et al.}, the entanglement length $N_e$ was calculated as 
	\begin{equation}
		\frac{1}{N_e} = \frac{d \langle Z \rangle}{d N_c}
	\end{equation}
	$\frac{d \langle Z \rangle}{d N_c}$ can be identified as the slope of a linear fit of $\langle Z \rangle$ \textit{vs} $N_c$.

	\subsection*{Calculation of further structural and dynamical properties}
	All structural and dynamical properties were calculated using an in-house implementation build on top of the MDAnalysis tool \cite{Michaud-Agrawal.2011} to handle large trajectories. The scripts are provided together with the ACELearner tool.

	\section*{Author contributions}
	M. Fischer: conceptualization, data curation, formal analysis, methodology, software, validation, visualization, writing - original draft.
	A. Heuer: supervision, writing - review \& editing.
	
	\section*{Conflicts of interest}
	There are no conflicts to declare.
	
	\section*{Data availability}
	The ACELearner software tool is provided via Github at \url{https://github.com/TibMont/ACElearner}.
	
	\section*{Acknowledgements}
	M.F. acknowledges funding from the BMBF (InterKI, 16DHBKI049).
	Parts of the MD simulations and analysis have been performed on the HPC cluster PALMA-II of the University of Münster.
	
	\bibliography{literature_complete}
	\bibliographystyle{unsrt}
	
	%%%%%%%%%%%%%%%%%%%%%%%%%%%%%%%%%%%%%%%%%%%%%%%%%%%%%%%%%%%%%%%%%
	% Start of SI
	
	\title{\LARGE{Supporting Information:} \\ \Large{From oligomers to entangled polymers: How to train a transferable machine learning interatomic potential}}
	
	\date{}
	
	\maketitle
	\setcounter{section}{0}
	\setcounter{page}{1}
	\setcounter{figure}{0}
	\setcounter{table}{0}
	
	\renewcommand{\thefigure}{S\arabic{figure}}
	\renewcommand{\thetable}{S\arabic{table}}
	\renewcommand{\theequation}{S\arabic{equation}}
	\renewcommand{\thesection}{S\arabic{section}}
	\renewcommand{\thepage}{S\arabic{page}}
	
	\renewcommand\floatpagefraction{0.99}
	\newpage
	
	\newpage
	\section{Pacemaker Settings}
	\label{si_sec:pacemaker_settings}
	
	\begin{table}[h!]
		\centering
		\caption{Settings for the \textit{pacemaker} tool.}
		\begin{tabular}{l| l l}
			\hline
			Property & Key & Value \\
			\hline
			Elements & elements & C, H \\
			Embedding & embeddings & \makecell[l]{Shifted Finnis-Sinclair type \\ (Linear ACE)} \\
			\hline
			Basis function type & radbase & Bessel \\
			Basis function parameter & radparameters & 5.25 \\
			Basis function outer cutoff & rcut & 7.0 $\mathrm{\AA}$ \\
			Basis function outer cutoff & dcut & 0.01 $\mathrm{\AA}$\\
			Basis function inner cutoff & r\_in & 0.8 $\mathrm{\AA}$\\
			Basis function inner cutoff & delta\_in & 0.5 $\mathrm{\AA}$\\
			Cutoff function for basis functions & NameOfCutoffFunction & cos \\
			$N_\mathrm{max}$ by order (1,2,3) & nradmax\_by\_orders & 40, 3, 2 \\
			$l_\mathrm{max}$ by order (1,2,3) & lmax\_by\_orders & 0, 2, 2 \\
			\hline
			Size of test set & test\_size & 0.05 \\
			Weighing of forces vs. energy ($\kappa$) & kappa & 0.95 (0.05 for refining) \\
			L1-regularization & L1\_coeff & $10^{-9}$ \\
			L2-regularization & L2\_coeff & $10^{-9}$ \\
			Optimizer & optimizer & BFGS \\
			Maximum number of optimizer steps & maxiter & 1000 \\
			\hline
			Weighting function for structures & type & EnergyBasedWeightingPolicy \\
			Maximum energy per structure & DEup & 2.0 eV\\
			Maximum total force per atom & DFup & 4.0 $\mathrm{eV\AA}$\\
			\hline
		\end{tabular}
		\label{tab:pacemaker_settings}
	\end{table}
	
	\newpage
	\section{Active Learning Settings}
	\label{si_sec:al_settings}
	\begin{table}[h!]
		\centering
		\caption{Settings of the active learning loop.}
		\begin{tabular}{l| l l}
			\hline
			Property & Key & Value \\
			\hline
			Maximum Iterations & max\_iterations & 21 \\
			\makecell[l]{Maximum test simulations\\ per iteration} & test\_simulations & 33 \\
			\makecell[l]{Number of extrapolated structures \\ added to training set per iteration} & n\_select\_al & 50 \\
			\hline
			\makecell[l]{Addition of molecules \\ at iterations} & iter\_add\_chain & 8, 12, 16 \\
			\makecell[l]{Number of molecules \\ after adding further ones} & n\_add\_chain & 2, 4, 6 \\
			Atom type to species mapping & mapping & C C H \\
			\hline
			Performing volume scan & perform\_volume\_scan & True \\
			Number of potentials fitted in final iteration & n\_final\_potentials & 6\\
			\hline
			Molecule for initial simulation (SMILES) & smiles & CCCCCCCC \\
			Box length initial simulation & box\_length & 50.1 $\mathrm{\AA}$ \\
			Output frequency initial simulation & freq\_outputs & 10000 \\
			Initial simulation time & n\_runs & 2000000 \\
			Temperature initial simulation & v\_temp & 430 \\
			Timestep initial simulation & timestep & 0.0005 ps \\
			Cutoff radius initial simulation & cutoff & 10 $\mathrm{\AA}$ \\
			\hline
			\makecell[l]{Molecules considered during \\further simulations} & simulation\_scan:smiles & \makecell[l]{CCCCCC, \\CCCCCCCC, \\CCCCCCCCCC} \\
			Box length test simulations & box\_length & 20.1 $\mathrm{\AA}$ \\
			Output frequency test simulations & freq\_outputs & 100 \\
			Test simulation time & n\_runs & 100000 \\
			Timestep test simulation & timestep & 0.0005 ps \\
			Cutoff radius test simulation & cutoff & 10 $\mathrm{\AA}$ \\
			Ensemble during equilibration & equil\_mode & nvt \\
			Ensemble during production & prod\_mode & nvt \\
			Calculation of extrapolation grade & freq\_outputs\_gamma & 1 \\
			Extrapolation grade for dumping structure & write\_gamma & 2.0 \\
			\makecell[l]{Extrapolation grade for \\terminating test simulation} & stop\_gamma & 3.5 \\
			\hline
			Mode for volume scans & volume\_scan\_mode & interval \\
			Performing of volume scans every $n$ iterations & volume\_scan\_steps & 1 \\
			\makecell[l]{Maximum number of structures \\considered for volume scan} & n\_structures & 25 \\
			Data points per volume scan & n\_data\_points & 6 \\
			Scaling factors for volume scan & scaling\_range & [1.01, 1.2] \\
			\hline
		\end{tabular}
		\label{tab:active_learning_settings}
	\end{table}
	
	\newpage
	\section{Dataset for descriptor comparison and RMSE calculation of forces and energies during AL}
	\label{si_sec:dataset_descriptor}
	
	For the comparison of different local atomic descriptors a set of structures was employed, which has been collected during AL. This ensures that the descriptor space is well sampled.
	AL has been performed for alkanes with chain lengths $N_c$ of 2, 3, 4, 8 and 16 over 20 iterations. After completion of all AL iterations, 50 structures (~5 \% of all structures collected during AL) for each $N_c$ were selected randomly. 
	
	A similar approach was employed to create a test dataset for the calculation of the RMSE of forces and energies to evaluate the AL process afterwards.

	\newpage
	\section{Data set for the refined ACE potential}
	
	\begin{table}[h]
		\centering
		\caption{Number of structures used for training the refined ACE potential. Unique structures refers to the number of unique structures included in the data set, excluding all duplicates with scaled volumes.}
		\begin{tabular}{l|c c c c}
			\hline
			$N_c$ & $N_\mathrm{mol}=1$ & $N_\mathrm{mol}=2$ & $N_\mathrm{mol}=4$ & $N_\mathrm{mol}=6$ \\
			\hline
			& \multicolumn{4}{c}{Unique} \\
			\hline
			6 & 151 & 64 & 32 & 24\\
			8 & 384 & 119 & 75 & 44\\
			10 & 306 & 118 & 150 & 139\\
			\hline 
			& \multicolumn{4}{c}{Volume scan duplicates included} \\
			\hline
			6 & 151 & 109 & 92 & 64\\
			8 & 384 & 294 & 225 & 134\\
			10 & 306 & 293 & 425 & 439\\
			\hline
		\end{tabular}
		\label{tab:data_set}
	\end{table}
	
	\newpage
	\section{Simulation Setup}
	
	\begin{table}[h]
		\centering
		\caption{Setup of simulations for different chain lengths $N_c$. The box size refers to the length of the equilibrated cubic simulation box for $T=430\:\mathrm{K}$ using the OPLS-AA force field.}
		\scriptsize
		\begin{tabular}{l|c c | c c c c c | c}
			\hline
			$N_c$ & $N_\mathrm{mol}$ & $N_\mathrm{atoms}$ &  \multicolumn{5}{c|}{Simulation time [ns]} & \makecell{Box size [$\mathrm{\AA}$] \\ Ref. 430 K} \\
			& & & - & \makecell{Equi 1. (NPT) \\  at 630 K \\ (Comp. 1000 bar)}  & \makecell{Cool down \\ (NPT) to $T_\mathrm{aim}$} & \makecell{Equi 2 (NPT) \\ at. $T_\mathrm{aim}$} & \makecell{Prod \\ (NPT)} &  \\
			\hline
			2 & 1019 & 8152 & & 0.5 & 0.1 & 0.1 & 10 & 321.0\\
			3 & 682 & 7502 & & 0.5 & 0.1 & 0.1 & 10 & 288.7\\
			4 & 512 & 7168 & & 0.5 & 0.1 & 0.1 & 10 & 256.6\\
			5 & 409 & 6953 & & 0.5 & 0.1 & 0.1 & 10 & 227.3\\
			6 & 341 & 6820 & & 0.5 & 0.1 & 0.1 & 10 & 201.7\\
			7 & 292 & 6716 & & 0.5 & 0.1 & 0.1 & 10 & 47.0\\
			8 & 256 & 6656 & & 0.5 & 0.1 & 0.1 & 10 & 45.7\\
			10 & 204 & 6528 & & 0.5 & 0.1 & 0.1 & 10 & 44.2 \\
			12 & 170 & 6460 & & 0.5 & 0.1 & 0.1 & 10 & 43.3\\
			14 & 146 & 6424& & 0.5 & 0.1 & 0.1 & 10 & 42.8\\
			16 & 128 & 6400 & & 0.5 & 0.1 & 0.1 & 10 & 42.5\\
			32 & 64 & 6272 & & 0.5 & 0.1 & 0.1 & 10 & 41.3\\
			64 & 32 & 6208 & & 0.5 & 0.1 & 0.1 & 10 & 40.7\\
			128 & 16 & 6176 & & 0.5 & 0.1 & 0.1 & 10 & 40.5\\
			256 & 8 & 6160 & & 0.5 & 0.1 & 0.1 & 10 & 40.3\\
			\hhline{=========}
			& & & \makecell{Equi. 0 (NVT) \\ at 630 K} & \makecell{Equi. 1 (NPT) \\ at 630 K \\ (Comp. 1000 / 10 bar)}  & \makecell{Equi. 2 (NPT)\\ at 630 K} & \makecell{Cool down \\ (NPT) to $T_\mathrm{aim}$} & \makecell{Prod \\ (NPT)}&  \\
			\hline
			512 & 25 & 38450 & 2 & 10+10 & 100 & 10 & 100 & 74.4\\
			\hline
		\end{tabular}
		\label{tab:sim_setup}
	\end{table}
	
	\newpage
	\section{Dependence of $E_\mathrm{pot}$ on carbon to hydrogen ratio}
	
	\begin{figure}[h!]
		\centering
		\includegraphics{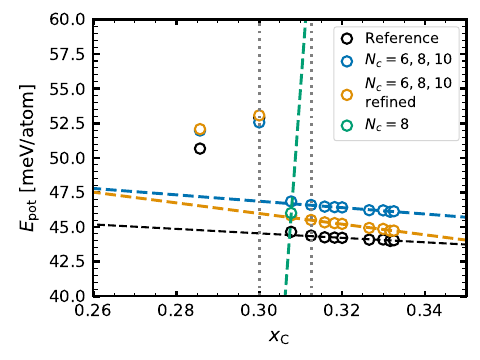}
		\caption{Dependence of $E_\mathrm{pot}$ on the ratio $x_\mathrm{C}$ of carbon to hydrogen atoms in a system, which depends on the chain length. The ratios corresponding to $N_c=6$ and $N_c=10$ are indicated by gray dotted lines. The dashed lines represent linear fits of $E_\mathrm{pot}$ \textit{vs.} $x_\mathrm{C}$ for systems in liquid phase ($N_c>6$).}
		\label{si_fig:e_pot_ratio}
	\end{figure}
	
	\newpage
	\section{Evaluation of the active learning process}
	
	\begin{figure}[h!]
		\centering
		\includegraphics[width=\textwidth]{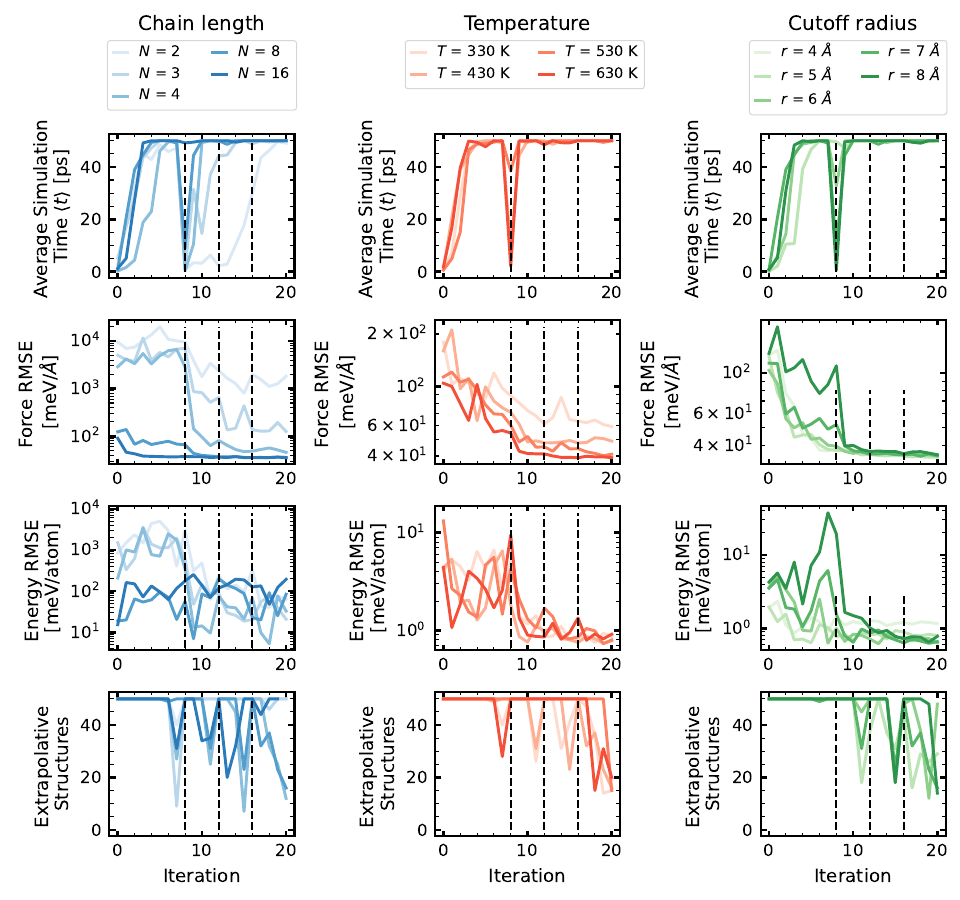}
		\caption{Evaluation of the active learning progress for varying chain lengths at 430 K and for varying temperature at a fixed chain length of 8. For both a ACE cutoff radius of $7\;\AA$ is applied. Furthermore, the cutoff radius is varied while the temperature is fixed to 430 K and the chain length to 8. The maximum average simulation time is 50 ps. Similar to Figure 3 in the main paper, RMSEs are averaged over a test data set which comprises for each setup structures from all conditions: Different chain lengths in the first column, different temperatures in the second column and different cutoff-radii in the third columnn.}
		\label{si_fig:active_learning}
	\end{figure}
	
	\newpage
	\section{Volume scan with non-refitted potentials}
	
	\begin{figure}[h!]
		\centering
		\includegraphics{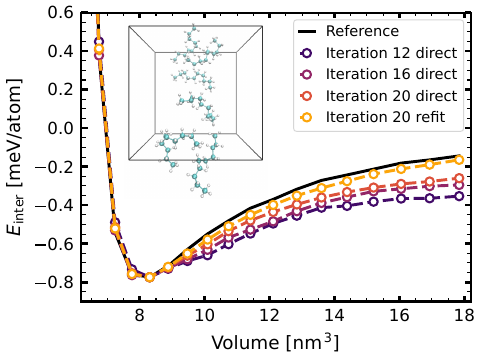}
		\caption{Inter-molecular potential energy for volume scans using non-refined  ACE potentials from different iterations of the active learning loop and the refined ACE potential after iteration 20. The reference is shown as a black line. All potentials are shifted such that their minima are aligned.}
		\label{si_fig:volume_scan_si}
	\end{figure}
	
	\newpage
	\section{Dissociation curves}
	
	\begin{figure}[h!]
		\centering
		\includegraphics[width=\textwidth]{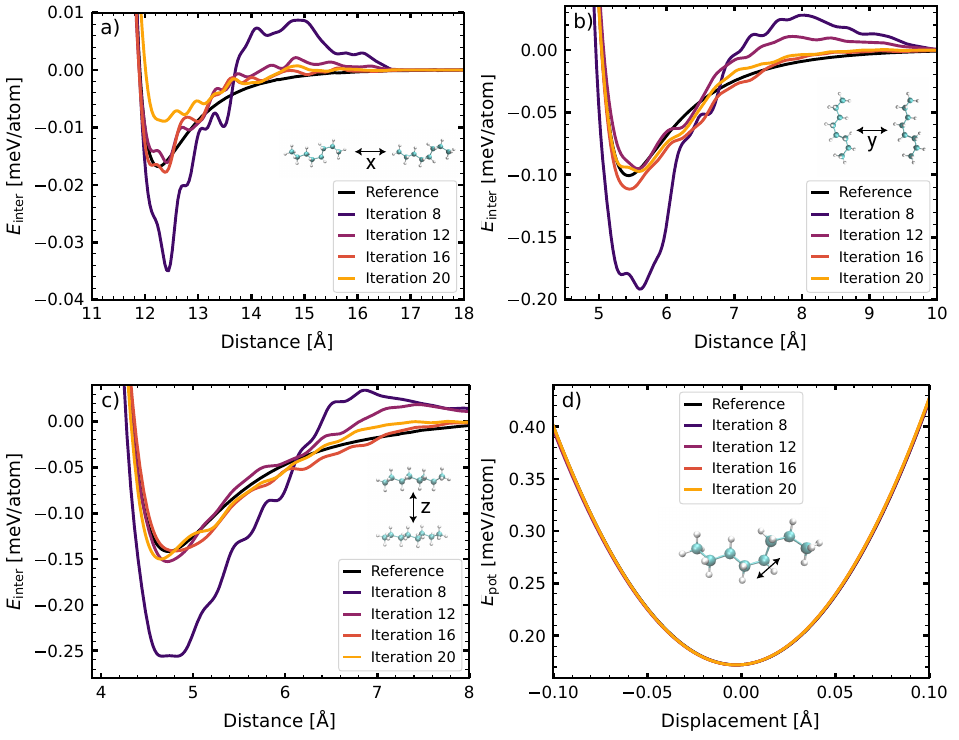}
		\caption{a-c) Dissociation curves of aligned molecules in a) x, b) y and c) z-direction. d) Systematic probing of bond distances by moving one central carbon atom while keeping all other atoms fixed. All curves were evaluated with the refined potentials after the given AL iteration.}
		\label{si_fig:dissociation}
	\end{figure}
	
	\newpage
	\section{Pairplots at different training stages}
	\label{si_sec:pairplots}
	
	The pairplots shown here were generated using the same test set as that employed for Figure 5 in the main paper, which includes clusters of up to six oligomers. Depending on the active learning iteration, the potentials were trained on progressively larger clusters, though some were exposed only to smaller clusters during their respective training stages.
	
	\begin{figure}[h!]
		\centering
		\includegraphics[width=0.8\textwidth]{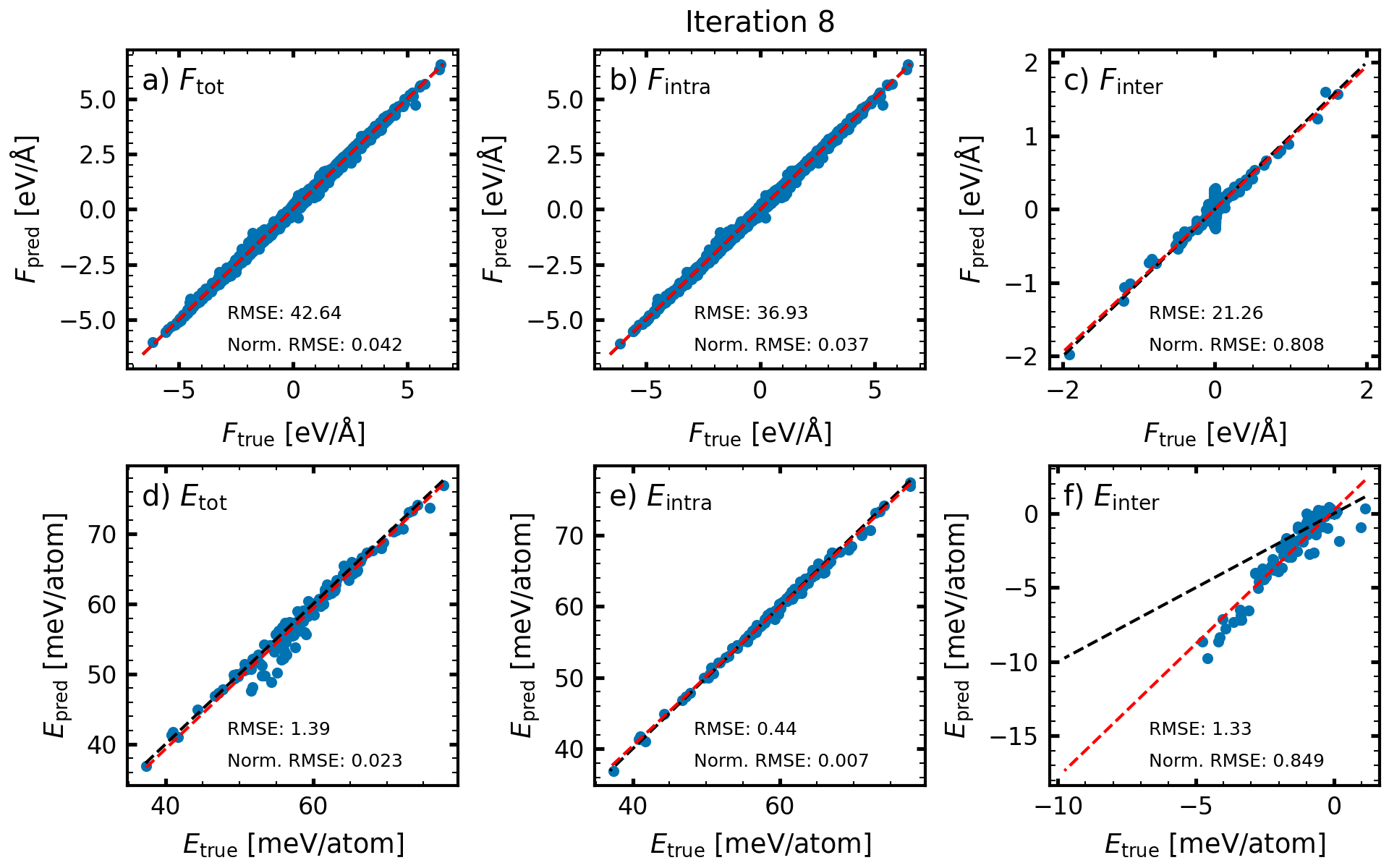}
		\caption{Pairplots of predicted versus true a) total forces, b) intramolecular forces, c) intermolecular forces, d) total energies, e) intramolecular energies and f) intermolecular energies for the ACE potential at AL iteration 8.}
		\label{fig:iteration_8}
	\end{figure}
	
	\begin{figure}[h!]
		\centering
		\includegraphics[width=0.8\textwidth]{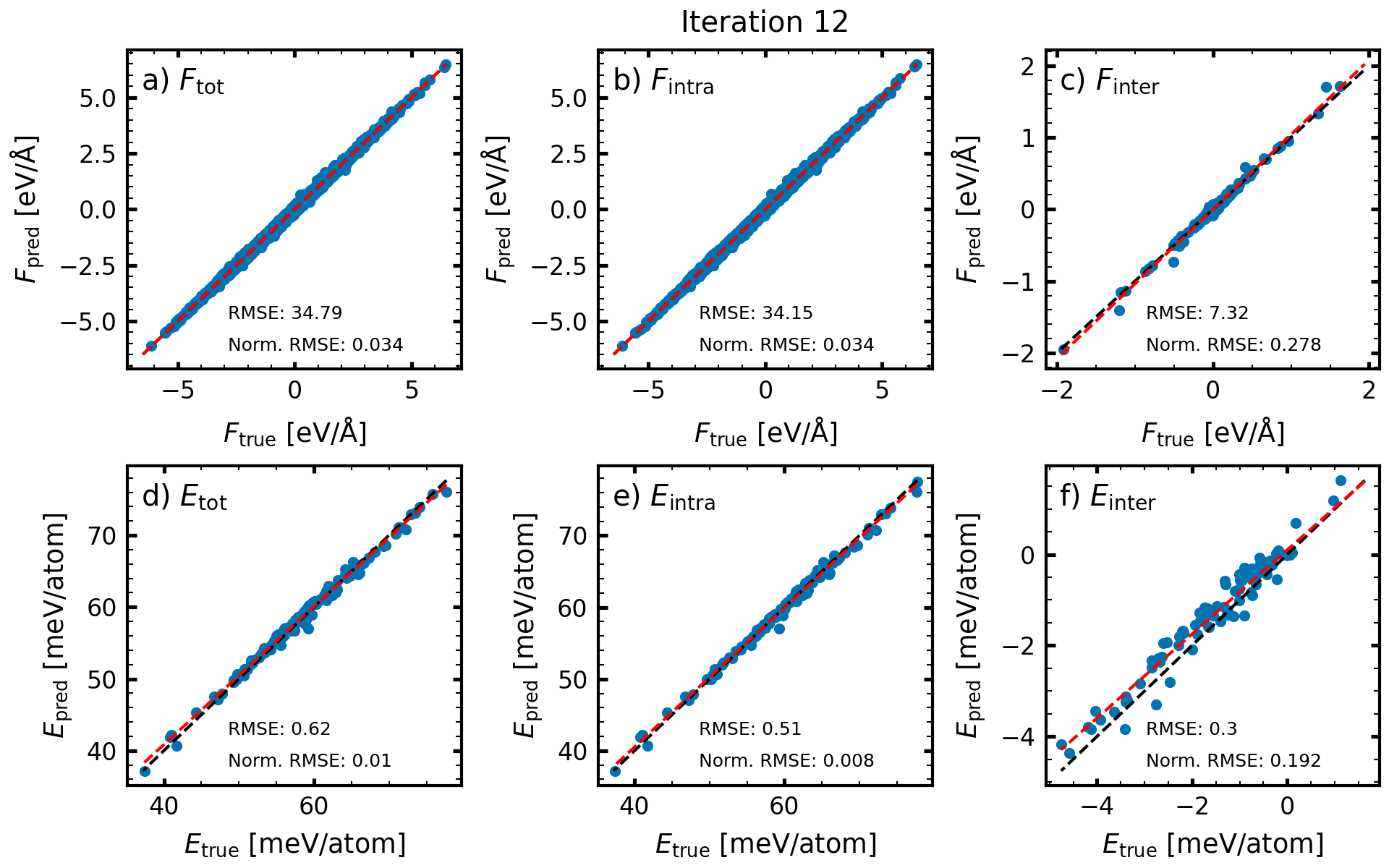}
		\caption{Pairplots of predicted versus true a) total forces, b) intramolecular forces, c) intermolecular forces, d) total energies, e) intramolecular energies and f) intermolecular energies for the ACE potential at AL iteration 12.}
		\label{fig:iteration_12}
	\end{figure}
	
	\begin{figure}[h!]
		\centering
		\includegraphics[width=0.8\textwidth]{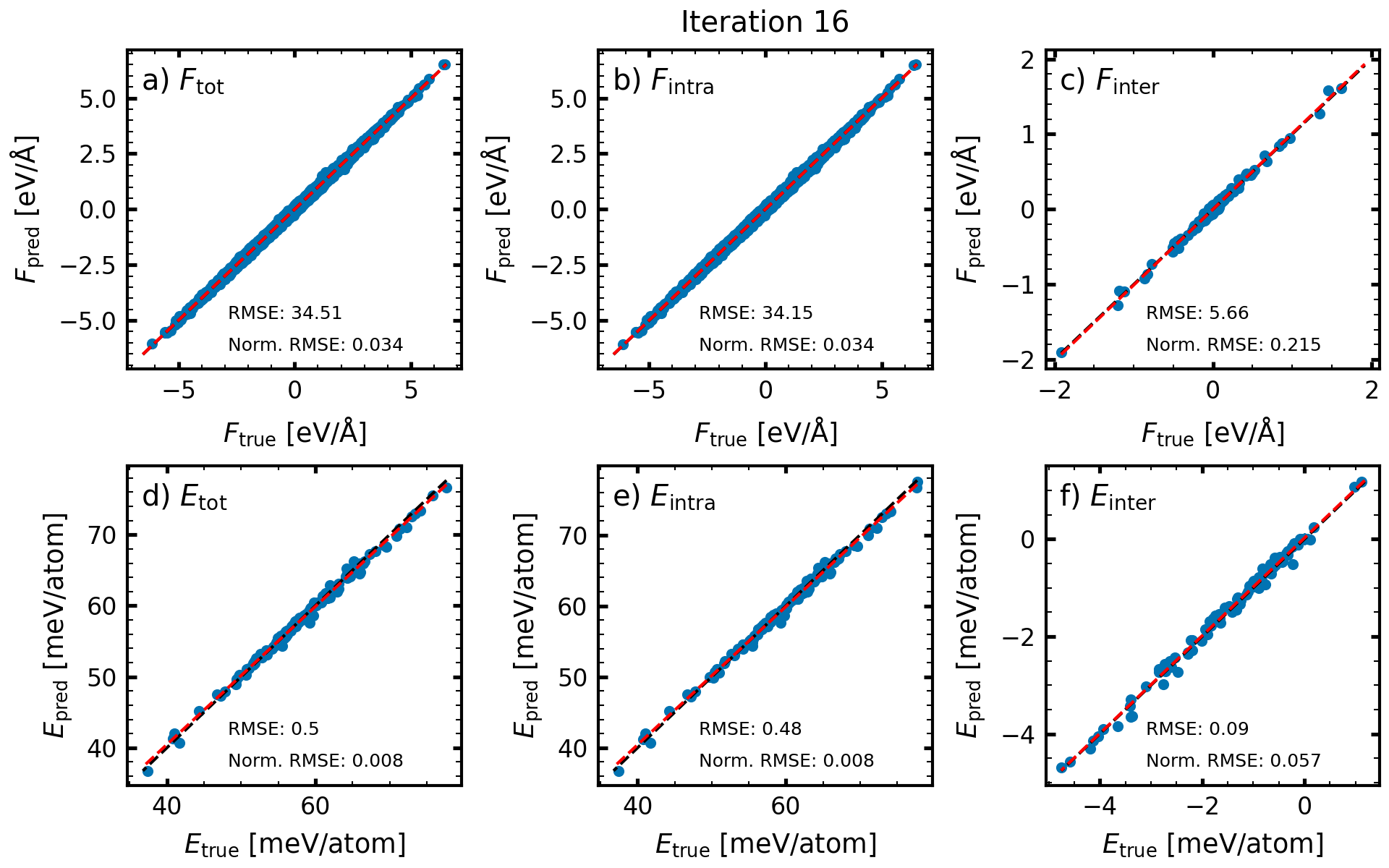}
		\caption{Pairplots of predicted versus true a) total forces, b) intramolecular forces, c) intermolecular forces, d) total energies, e) intramolecular energies and f) intermolecular energies for the ACE potential at AL iteration 16.}
		\label{fig:iteration_16}
	\end{figure}
	
	\begin{figure}[h!]
		\centering
		\includegraphics[width=\textwidth]{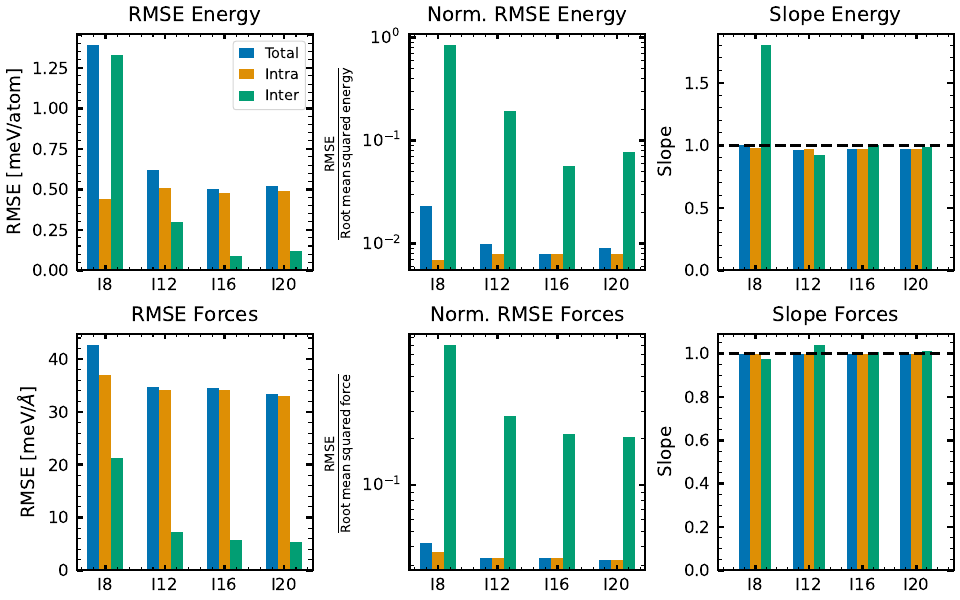}
		\caption{Improvement of the fitted ACE potentials during the active learning. Comparison of the RMSE and normalized RMSE of energy and forces after certain iterations and fitted slope of the predicted vs. true energy and forces. A slope of 1 indicates at least correct trends although their might be trivial systematic shifts in the energy.}
		\label{fig:al_improvement}
	\end{figure}

	\clearpage
	\newpage
	\section{Density}
	
	\begin{figure}[h!]
		\centering
		\includegraphics[width=0.9\textwidth]{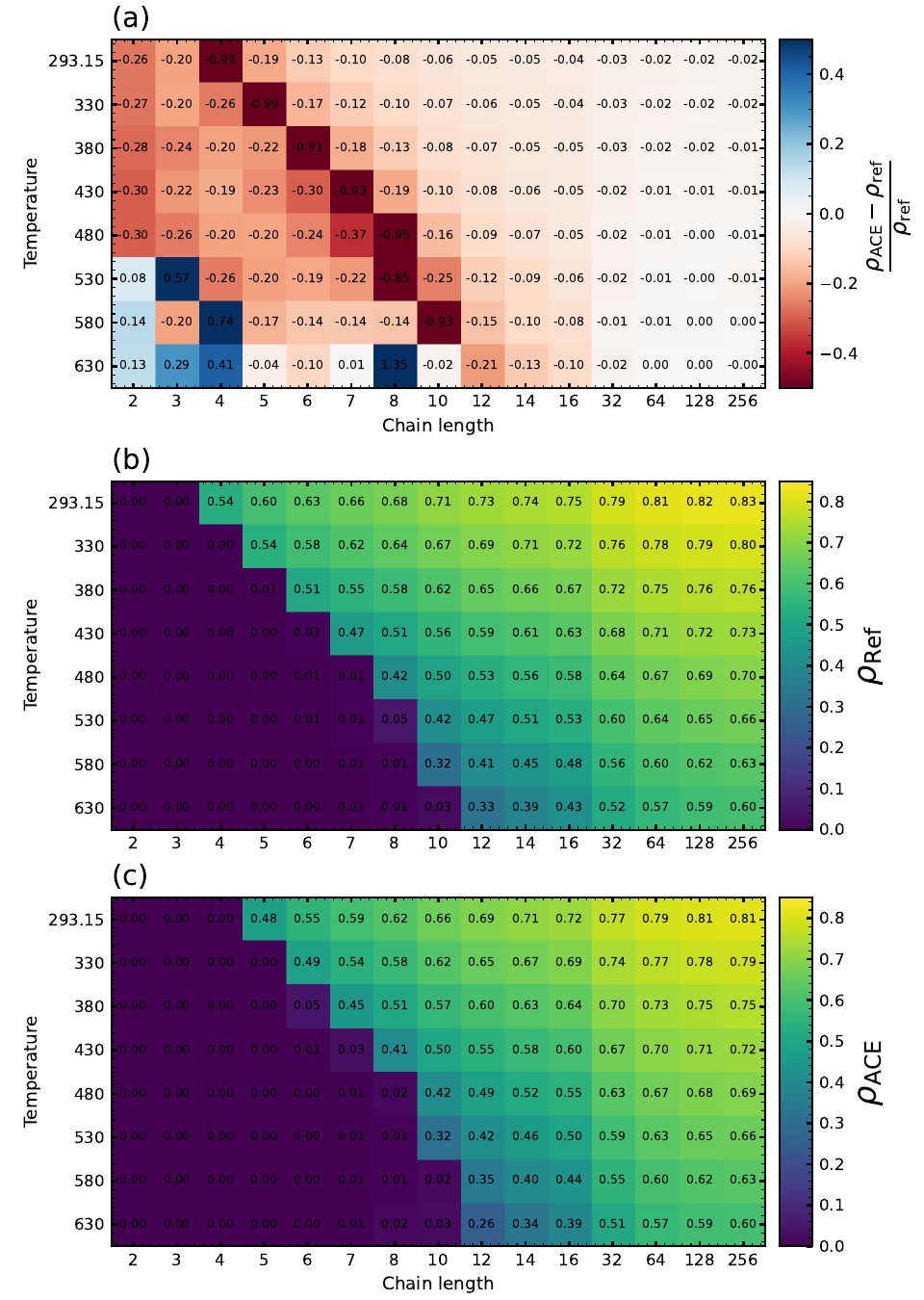}
		\caption{a) Relative difference in the densities between ACE and reference simulations. b) Density in reference simulations. c) Density in ACE simulations.}
		\label{fig:density_si}
	\end{figure}
	
	\newpage
	\section{Extrapolation grade and simulation time}
	\label{si_sec:sim_time}
	
	\begin{figure}[h!]
		\centering
		\includegraphics[width=0.9\textwidth]{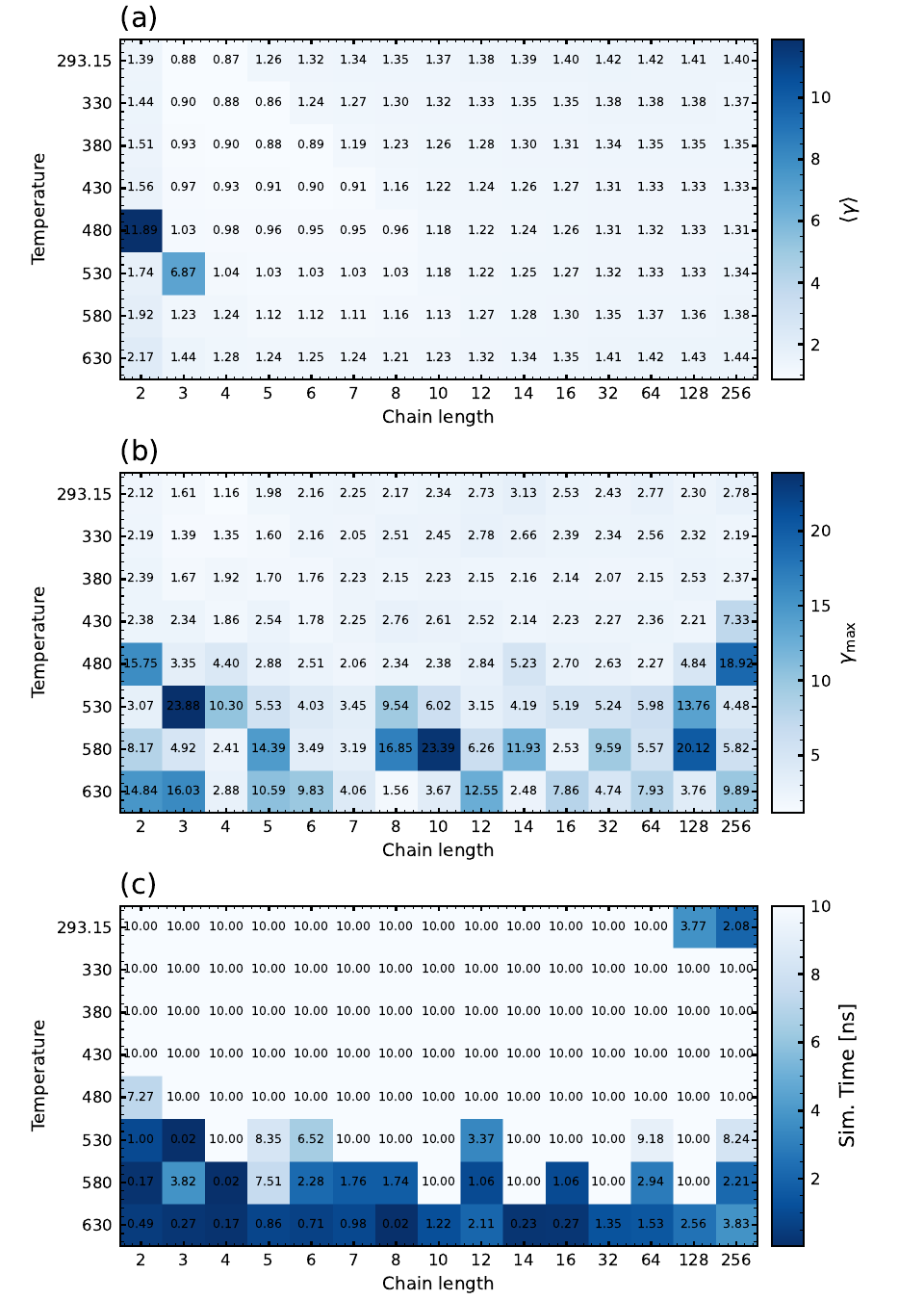}
		\caption{a) Average and b) maximum extrapolation grade $\gamma$ observed during the simulation before the stopping criterion with $\gamma_\mathrm{stop}>25$ was reached. c) Simulation time until $\gamma_\mathrm{stop}>25$ was reached. The maximum simulation time is 10 ns.}
		\label{fig:sim_time}
	\end{figure}
	
	\newpage
	\section{Bond distance, angle and dihedral distributions}
	
	\begin{figure}[h!]
		\centering
		\includegraphics[width=\textwidth]{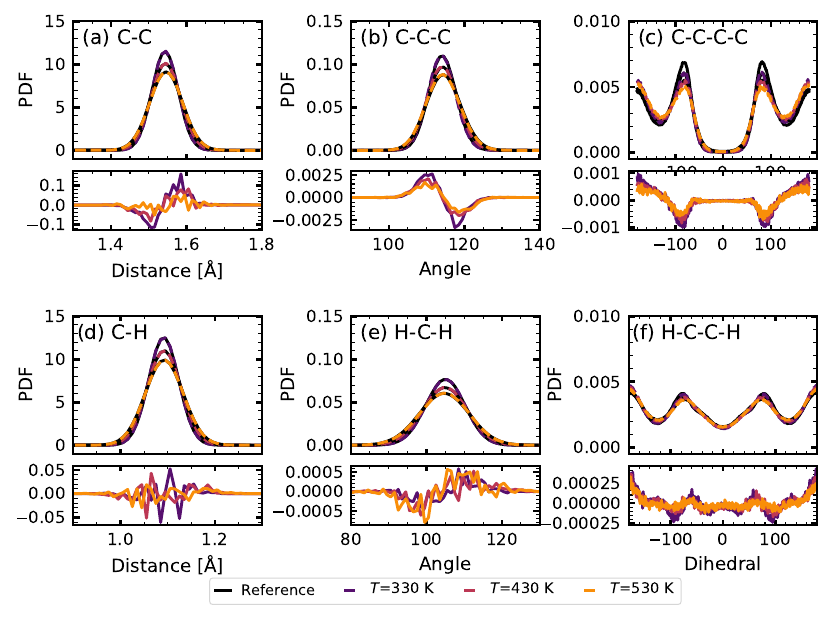}
		\caption{Comparison of bonds, angles and dihedral distributions obtained from simulations with the ACE potential with distributions obtained from the reference potential. a-c) Show distributions for carbon-backbone atoms only. d-f) show distributions including hydrogen atoms. The distributions are shown for $N_c=128$ and varying temperature.}
		\label{fig:structure_temp}
	\end{figure}
	
	\begin{figure}[h!]
		\centering
		\includegraphics[width=\textwidth]{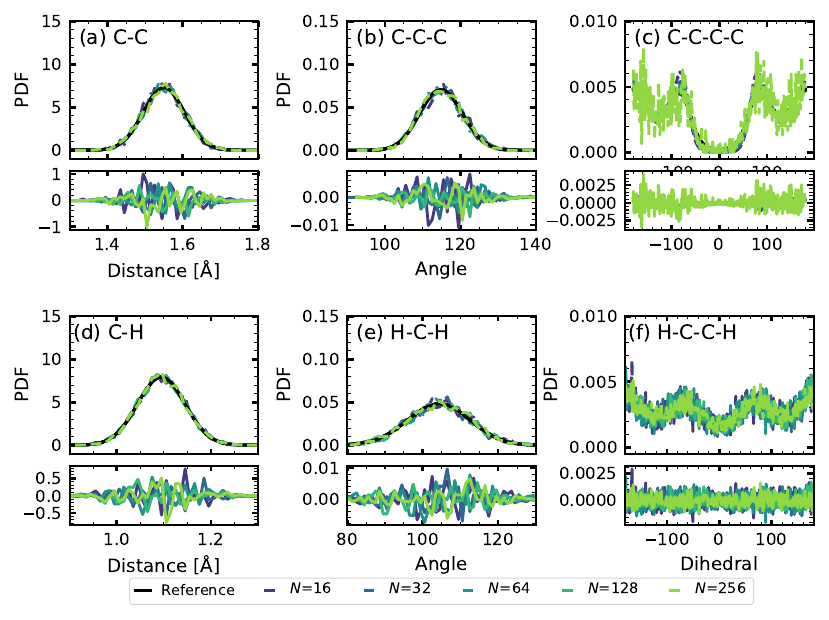}
		\caption{Comparison of bonds, angles and dihedral distributions obtained from simulations with the ACE potential with distributions obtained from the reference potential. a-c) Show distributions for carbon-backbone atoms only. d-f) show distributions including hydrogen atoms. The distributions are shown for $T=430\;\mathrm{K}$ and varying $N_c$.}
		\label{fig:structure_Ns}
	\end{figure}
	
	\clearpage
	
\end{document}